\documentclass[aps,preprint]{revtex4}
\usepackage{graphicx}
\usepackage{amsbsy}
\usepackage[thinspace,thinqspace]{SIunits}
\newcommand{\YRS}{YbRh$_2$Si$_2$}
\newcommand{\yrs}{YbRh$_2$Si$_2$\,}

\newcommand{\TN}{$T_{\rm N}$\,}

\newcommand{\kph}{$\kappa_{\rm ph}$\,}
\newcommand{\kelT}{$\kappa_{\rm el}(T)$\,}
\newcommand{\kphT}{$\kappa_{\rm ph}(T)$\,}

\newcommand{\et}{~\textit{et al.\,}}

\begin{document}

\begin{center}
{\large\bf Thermal and Electrical Transport across a
Magnetic Quantum Critical Point}\\[0.5cm]
Heike Pfau$^{1}$, Stefanie Hartmann$^{1,*}$, Ulrike
Stockert$^{1}$, Peijie Sun$^{1,**}$, Stefan Lausberg$^{1}$, Manuel
Brando$^{1}$, Sven Friedemann$^{1,+}$, Cornelius Krellner$^{1,+}$,
Christoph Geibel$^1$, Steffen Wirth$^{1}$, Stefan Kirchner$^{2,1}$,
Elihu Abrahams$^3$, Qimiao Si$^{4}$ and Frank Steglich$^1$\\
\textit{$^{1}$Max Planck Institute for Chemical Physics of Solids,
N{\"o}thnitzer Str.~40, 01187~Dresden, Germany}\\
{\em $^{2}$Max Planck Institute for the Physics of Complex Systems,
N{\"o}thnitzer Str.~38, 01187~Dresden, Germany}\\
\textit{$^{3}$Department of Physics and Astronomy, University of
California Los Angeles, 405 Hilgard Avenue, Los Angeles, CA 90095, USA}\\
\textit{$^{4}$Department of Physics and Astronomy, Rice University,
Houston, TX 77005, USA}\\
\textit{$^{*}$present address: Leibniz Institute for Solid State and
Materials Research, Helmholtz Str. 20, 01069~Dresden, Germany}\\
\textit{$^{**}$present address: Institute of Physics, Chinese
Academy of Sciences, Beijing 100190, China}\\
\textit{$^{+}$present address: Cavendish Laboratory, University of
Cambridge, J.J. Thompson Avenue, Cambridge, CB3 OHE, UK}\\
\end{center}
\vspace{0.5cm} {\bf A quantum critical point (QCP) arises at a
continuous transition between competing phases at zero
temperature. Collective excitations at magnetic QCPs give rise to
metallic properties that strongly deviate from the expectations of
Landau's Fermi liquid description\cite{Landau56}, the standard
theory of electron correlations in metals. Central to this theory
is the notion of quasiparticles, electronic excitations which
possess the quantum numbers of the bare electrons. Here we report
measurements of thermal and electrical transport across the
field-induced magnetic QCP in the heavy-fermion compound
\textbf{\YRS} \cite{Trovarelli00,Gegenwart02}. We show that the
ratio of the thermal to electrical conductivities at the
zero-temperature limit obeys the Wiedemann-Franz (WF) law above
the critical field, $B_c$. This is also expected at $B < B_c$,
where weak antiferromagnetic order and a Fermi liquid phase form
below 0.07 K ($B = 0$). However, at the critical field the
low-temperature electrical conductivity suggests a
non-Fermi-liquid ground state and exceeds the thermal conductivity
by about 10\%. This apparent violation of the WF law provides
evidence for an unconventional type of QCP at which the
fundamental concept of Landau quasiparticles breaks down
\cite{Si01,Coleman01,Senthil04}. These results imply that Landau
quasiparticles break up, and that the origin of this
disintegration is inelastic scattering associated with electronic
quantum critical fluctuations. Our finding brings new insights
into understanding deviations from Fermi-liquid behaviour
frequently observed in various classes of correlated materials.}

In metallic systems, continuously suppressing magnetic order gives
rise to a QCP \cite{Schofield10} and leads to non-Fermi-liquid
behaviour
\cite{Lohneysen94,Aronson95,Mathur98,Trovarelli00,Grigera.01}.
Whether quasiparticles persist near the QCP, however, is a
fundamental open issue. An established means to probe the fate of
the quasiparticles is to compare the thermal conductivity
($\kappa$) and electrical conductivity ($\sigma$). If
quasiparticles are well defined, the WF law specifies the
zero-temperature ($T = 0$) value of the Lorenz number $L \equiv $
$\kappa/T\sigma$ to be L$_0 =(\pi {\rm k_B})^2/3e^2$. Except for
superconductors \cite{Einstein22}, where the Lorenz ratio $L/{\rm
L}_0=0$, a violation of the WF law would constitute a direct
evidence for physics beyond the Fermi liquid theory. $L/{\rm L}_0$
becomes larger than one if there are additional carriers which
contribute to the heat current but not to the charge current
\cite{Wakeham11}. By contrast, $L/{\rm L}_0<1$ at $T = 0$ implies
a breakdown of Landau quasiparticles.

Heavy-fermion metals are prototype systems for antiferromagnetic
(AF) QCPs. These rare-earth or actinide-based intermetallics
contain both $f$-derived localised magnetic moments and itinerant
({\it spd}) conduction electrons, whose entanglement gives rise to
the Kondo effect and the concomitant composite quasiparticles
with huge effective mass. 
In these materials, two types of QCPs have been highlighted. The
spin-density-wave based theory relies on the fluctuations of the
AF order parameter \cite{Hertz76,Moriya85, Millis93}. Here, the
main part of the Fermi surface remains unaffected by the critical
fluctuations, leaving the quasiparticles intact. In a related
picture \cite{Wolfle11}, all states near the Fermi surface are
influenced by quantum critical fluctuations. At an unconventional
type of QCP, a breakdown of the Kondo entanglement disintegrates
all the heavy quasiparticles \cite{Si01,Coleman01,Senthil04}.
Neutron scattering and magnetic measurements \cite{Schroder00} in
CeCu$_{6-x}$Au$_x$, as well as de Haas-van Alphen
\cite{Shishido05} and thermodynamic and transport \cite{Park06}
measurements in CeRhIn$_5$ have been interpreted in terms of a
local Kondo-breakdown QCP \cite{Si01,Coleman01}. In \YRS, the weak
AF order is continuously suppressed by a tiny magnetic field
\cite{Trovarelli00,Gegenwart02,Custers03}. Electrical transport
and thermodynamic measurements have revealed multiple vanishing
energy scales \cite{Gegenwart08} and
a discontinuity of the Fermi surface across the QCP
\cite{Paschen04,Friedemann10}. These materials provide a setting
to characterize the quasiparticles near the AF QCP.

We focus on \YRS, in order to take advantage of the understanding
of its Fermi surface \cite{Paschen04,Friedemann10}. Fig.~1
displays the overall $T$-$B$ phase diagram (a) and the thermal
conductivity below 12 K (b). At $B = 0$, the compound orders
antiferromagnetically at $T_{\rm N}$ = 0.07 K. Increasing $B$ to
its critical value $B_{\rm c} \approx 0.06$ T ($\perp$ c)
 continuously suppresses $T_{\rm N}$ to zero, reaching the
QCP. Below $T_{\rm FL}$ the paramagnetic phase at $B > B_c$ is a
heavy Fermi liquid \cite{Gegenwart02}, in which the Fermi surface
is large as a result of the Kondo effect. However, in the AF phase
($B<B_c$), also a mass-enhanced Fermi liquid \cite{Gegenwart02},
the Fermi surface is small, without incorporating the
$f$-electrons \cite{Paschen04,Friedemann10}. The $T^*(B)$ line
defines a crossover of the Fermi surface as a function of the
control parameter $B$, and terminates at $B = B_c$ as $T
\rightarrow 0$. Upon cooling, the field range of quantum critical
behaviour shrinks to $B = B_c$ in the $T = 0$ limit, whereas a
Fermi-liquid ground state exists on either side of the QCP. The
QCP is clearly identified by an asymptotic ($T \rightarrow 0$)
linear temperature dependence of the electrical resistivity,
independent of sample quality \cite{Custers03,Gegenwart08}. In
addition, the width of the $T^*$ crossover is proportional to $T$,
extrapolating to a sharp jump of the Fermi surface at $T=0$
(Ref.~\onlinecite{Friedemann10}). While these measurements prove the
existence of two different states on either side of the QCP, they
left open the nature of not only the electronic excitations in the
quantum critical regime but also the dynamical processes
underlying the Kondo breakdown.

A previous study of the thermal and electrical transport in the
quasi-two-dimensional heavy-fermion metal CeCoIn$_5$
\cite{Tanatar07}, in which a magnetic QCP is suspected
\cite{Zaum11} but not identified, found that the WF law is
violated ($L/{\rm L}_0 \approx 0.8$ as $T \rightarrow 0$) for
$c$-axis transport but obeyed for in-plane transport. These
results were discussed in terms of putative strongly anisotropic
critical fluctuations, although how spin fluctuations may
invalidate the WF law was left as an open question. Combined
thermal and electrical transport has also been studied near the
QCP of ZrZn$_2$, which is considered a canonical
ferromagnetic-spin-fluctuation system \cite{Smith08}. Although the
two transport quantities in ZrZn$_2$ have different temperature
dependencies, with $L/$L$_0<1$, their extrapolated $T=0$ limits
satisfy the WF law.

\YRS\ provides a unique opportunity to study the fate of Landau
quasiparticles at QCPs beyond the spin-fluctuation description
and, likewise, the nature of the critical fluctuations associated
with the Kondo breakdown. The compound is also advantageous
because superconductivity is absent down to at least 0.01 K
(Ref.~\cite{Custers03}), unlike the case of CeCoIn$_5$. This not
only exposes the properties in the immediate vicinity of the AF
QCP but also facilitates the characterization of the
quasiparticles through the Lorenz ratio. \YRS\ is a magnetically
anisotropic metal; the possibility of quasi-two-dimensional
transport necessitates the usage of in-plane transport to probe
any quasiparticle breakdown \cite{McKenzie08}. The present study
will therefore focus on the thermal and electrical transport
within the tetragonal plane.

The thermal conductivity $\kappa(T)$ was measured between 0.025\,K
and 12\,K and is shown in Fig.~1b for $B = 0$. For comparison,
$\kappa_{\rm {WF}} (T) \equiv$ L$_0 T /\rho (T)$ was calculated
from the measured electrical resistivity $\rho (T)$ and is also
presented. Above 4 K, $\kappa (T)$ exceeds $\kappa_{\rm {WF}}(T)$
due to the contribution of phonons to the heat transport,
$\kappa_{\rm {ph}}(T)$, see Supplementary Information (SI). Below
4 K, $\kappa_{\rm {ph}}(T)$ is suppressed, and $\kappa (T)$
becomes smaller than $\kappa_{\rm {WF}} (T)$ down to about
0.035\,K and somewhat larger at even lower temperatures (Fig.~1a).

In order to investigate the WF law, we extrapolate the Lorenz
ratio $L(T)/{\rm L}_0 \equiv \rho(T)/w(T)$ to $T = 0$. Because a
QCP is a singular point in the phase diagram, and given that there
are temperature scales that vary as a function of the control
parameter and vanish at the QCP, the combination of isofield and
isothermal scans is crucial for the extrapolation (Sec.~VI, SI).

Figure 2 depicts the low-temperature behaviour of both the
electrical resistivity $\rho(T)$ and thermal resistivity $w(T) =
{\rm L}_0 T /\kappa (T)$ at zero field, $B=0.06$ T $\approx B_c$
and $B > B_c$. Here $w (T)$ has the same unit as $\rho(T)$.
Similar results at other magnetic fields are given in Fig.~S4 of
SI. This comparison shows that $w(T)$ exceeds $\rho(T)$ over a
wide range of temperature and field. Figures 3a,b and c,d display,
respectively, the difference $w(T) - \rho(T)$ and the Lorenz ratio
for the data shown in Figs.~2a--d. Corresponding plots for the
data shown in Fig.~S4 are presented in Figs.~S5a,b and c,d.

Below $T = 0.15$ K, at $B \geq 0.6$ T, $w(T) = \rho (T)$ within
the experimental resolution. This is illustrated for $B = 1$\,T in
Fig.~3b which shows that $w(T) -\rho(T)$ approaches zero in this
range of $T$ and $B$, and in Fig.~3d, which demonstrates that
$L(T)/{\rm L}_0 = 1$ within the experimental error. In this
high-field range, both $\Delta \rho(T) = [\rho(T)-\rho_0] \propto
T^2$ and $[w(T) - w_0] \propto T^2$ below the Fermi-liquid
crossover temperature, marked by arrows in Fig.~2 and Fig.~S4.
Here, $\rho_0$ and $w_0$ are the residual ($T \rightarrow 0$)
electrical and thermal resistivities, which are identical within
about 1\%. These results establish the validity of the WF law in
the Fermi-liquid phase for $B \ge 0.6$ T. For 0.2 T $\le B \le$
0.6 T, the results shown in Figs.~2c and S4d--f suggest similar
Fermi-liquid behaviour at lower temperatures.

The system is in the quantum critical regime \cite{Custers03} at
$B = 0$ and $T \gtrsim 0.1$ K, where $w(T) > \rho(T)$. Both
$\rho(T)$ and $w(T)$ decrease linearly with temperature below
about $0.3$ K which allows extrapolating the quantum critical
behaviour of $\rho(T)$ and $w(T)$ to the $T=0$ limit, giving $w_0
> \rho_0$. This is shown in Fig.~3a (dashed line) for
$w(T)-\rho(T)$, which is about 0.17 $\mu \ohm \,$cm at $T = 0$,
and in Fig.~3c for $L(T)/{\rm L}_0$, where the corresponding
extrapolation (dashed line) yields about $0.9$. Our extrapolation
follows path C$_1$ (Fig.~S8a, SI).

Upon cooling in zero field, $\rho (T)$ drops at the N\'eel
transition \cite{Gegenwart02} into the AF phase, reflecting the
freezing out of spin-disorder scattering. Below $T_{\rm N}$, $\rho
(T)$ exhibits a Fermi-liquid $T^2$ dependence
(Ref.~\cite{Gegenwart02}). We observe a drop in $w (T)$ as well;
it starts already at $T \approx 0.1$ K (Fig.~2a). Furthermore, we
find that $w(T) - \rho(T)$ below $T =$ 0.033\,K becomes negative
(Fig.~3a), and $L/{\rm L}_0$ exceeds one (Fig.~3c). An upturn of
$L(T)/{\rm L}_0$ just below $T_N$ was also observed in
Ref.~\cite{Tomokuni11} down to 0.05\,K, where $L(T)/{\rm L}_0$ is
still below one but appears to extrapolate to one as $T
\rightarrow 0$. Our observation of $L(T)/{\rm L}_0 > 1$ below $T=$
0.033\,K clearly shows that the thermal transport is not entirely
due to electronic-quasiparticle transport, as concluded in
Ref.~\cite{Tomokuni11}. Instead, the Fermi-liquid-type electronic
transport is masked by that of an additional heat channel. In the
SI we demonstrate that the additional thermal conductivity is due
to AF magnons; this magnon contribution will vanish in the $T=0$
limit, as is inferred from the specific-heat data \cite{Custers03}
measured down to 0.018\,K (see SI). Therefore, at $B = 0$ the WF
law is expected to hold in the $T=0$ limit.

At $B = 0.06$\,T $\approx B_c$, $\rho(T)$ is linear below 0.12\,K
down to the lowest measured temperature, as is $w(T)$ below about
0.2\,K (Fig.~2b). At $T \approx$ 0.07\,K, $w(T)$ shows a downturn
which is similar to, though considerably weaker than, that at $B =
0$ which sets in at higher temperature
(Fig.~\ref{fig:rho_and_w}a). We interpret this feature as the
contribution of overdamped magnons in the paramagnetic regime
above the reduced $T_{\rm N}$, see SI; as in the case of $B = 0$,
this magnetic contribution is expected to vanish in the $T
\rightarrow 0$ limit. Extrapolating the $T$-linear electrical
resistivity and the \textit{electronic} thermal resistivity, which
is also linear in $T$ between 0.07 -- 0.2\,K, to $T=0$ we find
$(w_0 - \rho_0) > 0$ and $L(T\!\rightarrow \! 0)/{\rm L}_0 < 1$,
similar to the behaviour at $B = 0$ (Figs.~3a, c). Here, our
extrapolation is taken near the path C (Fig.~S8a, SI).

These results provide an overall picture that can be placed in the
context of the phase diagram of Fig.~1a. For fields sufficiently
above the critical field $B_c$, the WF law is obeyed at low
temperatures. At the same time, the data at $B=0$ can be
interpreted as validating the WF law in the $T=0$ limit, {\it
i.e.}, in the AF ground state. The validity of the WF law at
magnetic fields away from $B_c$ and for sufficiently low
temperatures is consistent with a field-induced continuous quantum
phase transition between two Fermi liquids with, respectively,
small and large Fermi surfaces, which has been inferred from
magnetotransport and thermodynamic measurements
\cite{Gegenwart08,Paschen04,Friedemann10}. In contrast, the data
in the paramagnetic quantum critical regime are extrapolated to a
$T=0$ limit that violates the WF law.

The isothermal field dependence $L(B)/{\rm L}_0$ further clarifies
these results. This is given in Fig.~\ref{fig:w_vs_T2}e, which
shows a shallow minimum near a field that tracks the $T^{*}(B)$
line in Fig.~\ref{fig:YRS_PD}a. The minimum narrows as temperature
is reduced and extrapolates, as $T \rightarrow 0$, to an abrupt
dip at $B = B_c$ (see SI); the extrapolated $T=0$ value at that
point is about $0.9$ ({\it cf.} Fig.~3c). The systematic evolution
of $L/{\rm L}_0$ versus $B$ and $T$ provides evidence for the
intrinsic nature of the apparent violation of the WF law in \YRS.

Our findings shed considerable new light on the dynamical
electronic processes occurring at the QCP. Quasiparticles
disintegrate at a Kondo-breakdown QCP, as illustrated in Fig.~4.
The large Fermi surface incorporates both the conduction electrons
and delocalized $f$-electrons, while the small Fermi surface
involves only the conduction electrons. Because the quantum phase
transition is continuous, this change of the Fermi surface must
result from inelastic processes that operate near the QCP. Such
dynamical processes must be electronic,
extending to zero energy when the system is precisely at the QCP.
Correspondingly, the quasiparticle residue of the large Fermi
surface, $Z_{\rm L}$, and that of the small Fermi surface, $Z_{\rm
S}$, must reach zero as the QCP is approached from the
paramagnetic and AF sides, respectively. At the critical value of
the control parameter, they satisfy dynamical scaling:
\begin{eqnarray}
Z_{\rm L} (\pmb{k}^{\rm L}_{\rm F}, T, \omega) = a_{\rm L}
T^{\alpha}
\varphi_{\rm L} (\omega/T), \nonumber \\
Z_{\rm S} (\pmb{k}^{\rm S}_{\rm F}, T,\omega) = a_{\rm S}
T^{\beta} \varphi_{\rm S} (\omega/T) . \label{z_l_s}
\end{eqnarray}
These scaling forms of $Z_{\rm L}$ and $Z_{\rm S}$ capture the
physics of the critical Kondo breakdown. The latter arises from
the dynamical competition between RKKY and Kondo interactions
which, respectively, promote small and large Fermi surfaces. The
resulting critical fluctuations between the small and large Fermi
surfaces amount to quantum critical inelastic scatterings for the
electronic heat carriers, which lead to $L/{\rm L}_0 <1$ in the
quantum-critical region. The vanishing quasiparticle weights,
$Z_{\rm L}$ and $Z_{\rm S}$, imply that such quantum fluctuations
and the concomitant fluctuating Fermi surfaces persist at the QCP,
thereby making it natural for $L/{\rm L}_0 < 1$ even in the $T=0$
limit. In this way, our observation provides evidence for
electronic quantum critical fluctuations which are naturally
associated with the abrupt reconstruction of the Fermi surface.

Our study is to be contrasted with those that feature extra
charge-neutral fermionic heat carriers. For example, over the
intermediate temperature range that corresponds to the
quasi-one-dimensional case of the ``purple bronze''
Li$_{0.9}$Mo$_6$O$_{17}$, one expects spin-charge separation. The
chargeless spinons, which contribute to the heat conductivity,
should give rise to $L/{\rm L_0} >1$, and this is indeed observed
\cite{Wakeham11}. In our case, the non-Fermi-liquid excitations
carry both charge and heat currents and are subject to inelastic
scatterings from quantum critical electronic fluctuations. As
discussed in Sec.\ VI, SI, this leads to a Lorenz ratio less than
one. Because it reflects the different degrees to which
non-Umklapp processes contribute to the electrical and heat
resistivities, the deviation in this case is expected to be more
modest; this is consistent with the 10\% effect observed here.

Our results indicate that a breakdown of Landau quasiparticles
accompanies the vanishing of a quantum critical energy scale,
$T^*(B)$, in \YRS. This linkage between the emergence of
non-Fermi-liquid excitations and vanishing of energy scales
provides a tantalizing connection between quantum critical
heavy-fermion metals and the high-$T_{\rm c}$ cuprates near
optimal doping, where the pseudogap energy scale collapses and
Fermi-liquid quasiparticles are destroyed over the entire Fermi
surface \cite{Casey08}.

\vspace*{0.8cm}\noindent{\bf \large Methods} \newline The samples
used in this work belong to the same batch and have been well
characterized previously \cite{Gegenwart02,Custers03}. Thermal and
electrical transport coefficients were obtained from the same
rectangular-shaped ($4.2 \times 0.5 \times 0.1$ mm$^{3}$) single
crystal (sample 1) with the same contact geometry. This allows a
reliable determination of the Lorenz ratio $L(T)/{\rm
L}_0=\rho(T)/w(T)$, since the geometry factor $l/A$ cancels out,
where $l$ and $A$ are the length and the cross-section of the
sample, respectively ({\it cf.} SI). Additional measurements of
the electrical resistivity were performed on a second single
crystal from the same batch, but with different geometry factor
(sample 2). As described in SI, the measured resistivity values
could perfectly be rescaled by a factor $1.25 \pm 0.03$ and
corrected by a difference in residual resistivity of
0.22\,$\mu\Omega$cm. Heat and charge currents as well as the
magnetic field were applied within the basal tetragonal plane.
However, we did not consider the distinction between the [100] and
the [110] directions within the basal plane. The parallel
orientation of the magnetic field, supplied by a superconducting
solenoid, to the heat and charge flow allows to neglect the
contributions of transverse effects (Nernst and electrical/thermal
Hall effects) in all measurements.

\vspace*{1.8cm} \noindent{\bf Supplementary Information}
accompanies the paper on {\bf www.nature.com/nature}.

\vspace{0.5cm}

\noindent{\bf Acknowledgements} We thank P. Coleman, R. Daou, P.
Gegenwart, N. E. Hussey, K. Ingersent, G. Kotliar, A. P.
Mackenzie, H. von L\"ohneysen, J. Schmalian, A. J. Schofield, T.
Senthil, S. Shastry, and Z. Te\'sanovic for useful discussions.
The work has in part been supported by the DFG Research Unit 960
``Quantum Phase Transitions'', NSF Grant DMR-1006985 and the
Robert A.\ Welch Foundation Grant No.\ C-1411. E. A., S. K., Q. S.
and F. S. acknowledge the support in part by the National Science
Foundation under Grant No. 1066293 and the hospitality of the
Aspen Center for Physics.

\vspace{0.5cm}

\noindent{\bf Author contributions} All authors contributed
substantially to this work.

\vspace{0.5cm}

\noindent{\bf Author information} Reprints and permissions
information is available at www.nature.com/reprints. The authors
declare no competing financial interests. Readers are welcome to
comment on the online version of this article at
www.nature.com/nature. Correspondence and requests for materials
should be addressed to F. S. (steglich@cpfs.mpg.de).

\newpage

\begin{figure}[ht]
\caption{\label{fig:YRS_PD} \textbf{Phase diagram and thermal
conductivity of YbRh$_2$Si$_2$}. \textbf{a},
Temperature-magnetic-field phase diagram, indicating  the
antiferromagnetic phase (AF) boundary ($T_{\rm N}$, solid line)
and the crossovers between non-Fermi-liquid and Fermi-liquid (FL)
regimes ($T_{\rm FL}$, dashed line) as well as between small and
large Fermi surfaces ($T^*$, double-dashed line). The crossover
width at $T^*$ is proportional to temperature (shaded region)
(from \cite{Friedemann10}). The magnetic field, $B$, was applied
within the basal tetragonal, easy magnetic plane, $\perp\! c$.
Arrows indicate fields at which combined thermal and electrical
transport measurements were performed
(Figs.~\ref{fig:rho_and_w}a-c). The WF law is strictly
defined only in the $T = 0$ limit and is expected to describe the
electronic transport of a Fermi liquid. This is illustrated in the
low-$T$ transport properties of the field-induced paramagnetic
phase, $B > B_c$ (Figs.~\ref{fig:rho_and_w}c,d). It is
also expected in the AF phase, $B < B_c$: Here, at finite
temperature the electronic thermal conductivity, $\kappa_{el}$, is
masked by a contribution due to magnons, $\kappa_m$ (see text).
However, as $T \rightarrow 0$, $\kappa_m$ vanishes such that the
heat transport is purely electronic, and the WF law is valid.
\textbf{b}, Thermal conductivity, $\kappa$, as a function of
temperature, $T$, at zero field. The solid line displaying
$\kappa_{\rm WF}(T)= {\rm L}_0 T/\rho (T)$ was obtained under the
assumption of the WF law to hold in the whole range of
temperatures $T \le$ 12 K; here, $\rho (T)$ is the electrical
resistivity and ${\rm L}_0 = 1/3 (\pi \rm k_{B}/e)^2$ Sommerfeld's
constant. The dashed line shows the phonon contribution
$\kappa_{\rm ph}(T)$, as discussed in the Supplementary
Information. Inset: same data below $T = 0.1$\,K.}
\end{figure}

\begin{figure}[ht]
\caption{\label{fig:rho_and_w} \textbf{Thermal and electrical
resistivity curves at low temperatures}. Thermal resistivity $w
(T) = {\rm L}_0 T/ \kappa (T)$ and electrical resistivity $\rho
(T)$ below $T = 0.5$ K for $B=0$ \textbf{a}, 0.06 T \textbf{b},
0.3 T \textbf{c} and 1 T \textbf{d}, $B \perp c$. Arrows in
\textbf{c} and \textbf{d} indicate the crossover to Fermi-liquid
behaviour (from Fig.~1a). Because $T_{\rm N}$ is very low
(0.07\,K), it is an experimental challenge to elucidate the
intrinsic behaviour of the thermal transport in the AF regime. We
have therefore made special efforts to not only extend the
heat-transport measurements at $B=0$ down to temperatures as low
as 0.025\,K, but also to reduce the statistical error of the data
by performing substantially more temperature scans than at finite
fields. The extrapolation specified by the dashed lines in
\textbf{a} and \textbf{b} corresponds to the trajectory C$_1$ and
one close to C, respectively, of Fig.~S8a, SI. Representative
error bars reflecting the standard deviation are shown for a few
selected temperatures.}
\end{figure}

\begin{figure}[ht]
\caption{\label{fig:w_vs_T2} \textbf{Violation/validity of the
Wiedemann-Franz law at $B \approx B_{\rm c}$}/$B > B_{\rm c}$.
\textbf{a}, Difference ($w-\rho$) vs $T$ at $B = 0$ and 0.06 T as
well as \textbf{b} 0.3 T and 1 T. \textbf{c, d}, Lorenz ratio
$L/{\rm L}_0 = \rho/w$ vs $T$ for the same fields as in \textbf{a,
b}. Within the experimental uncertainty, the validity of the
Wiedemann-Franz law is found below $ T \approx 0.15$ K for $B = 1$
T and is anticipated at lower temperature for $B \ge 0.3 $ T
\textbf{b, d}. Dashed lines in \textbf{a, c} indicate $T$-linear
behaviour of both $w(T)$ and $\rho (T)$ in the paramagnetic
non-Fermi-liquid regime. For $B = 0$, the onset of the deviation
from quantum critical behaviour (dashed lines in \textbf{a} and
\textbf{c}) seems to occur at about 0.07\,K, almost exactly where
the corresponding feature becomes visible at $B = 0.06$\,T, too.
This is in striking contrast to Fig.~\ref{fig:rho_and_w}a, showing
that the deviation in the $w(T)$ data for $B = 0$ sets in already
at $T \approx 0.1$\,K. The reason for this seeming discrepancy
lies in the pronounced drop of the electrical resistivity at
$T_{N} = 0.07$\,K (Fig.~\ref{fig:rho_and_w}a). The extrapolation
of the dashed lines in \textbf{a} and \textbf{c} to $T = 0$
demonstrates a violation of the Wiedemann-Franz law in a putative
paramagnetic, non-Fermi-liquid ground state. This ground state is
realized \cite{Gegenwart08} exactly at the critical magnetic field
$B_{\rm c}$, {\it cf.} Fig.\ \ref{fig:YRS_PD}a. Error bars are
derived from the standard deviation of the data in
Fig.~\ref{fig:rho_and_w}. \textbf{e}, Evolution of a shallow
minimum in the isothermal (0.1 K $\le T \le$ 0.3 K) $L(B)/{\rm
L}_0$ dependence. Data at lower $T$ are not included because of
the additional magnon heat transport at $B < B_c$ which will
vanish as $T \rightarrow 0$. These minima are related to the
$T^{*}(B)$ line of Fig.~\ref{fig:YRS_PD}a, {\it cf.} the crossover
fields (arrows) and widths (horizontal bars). Above $B^{*}(T) =
B(T^{*})$, $L/{\rm L}_0$ values are consistent with $L_{el}/{\rm
L}_0$ values in Fig.~S10 implying in the $T=0$ limit $L/{\rm L}_0
= 1$ at $B \neq B_c$ and an abrupt dip at $B = B_c$. Error bars as
in \textbf{c} and \textbf{d}.}
\end{figure}

\begin{figure}[ht]
\centering \caption{ \label{fig:z_vs_delta} \textbf{ The evolution
of the quasiparticle weights across a Kondo-breakdown quantum
critical point.} $Z_{\rm L}$, the quasiparticle residue at the
generic part (away from the ``hot spots", {\it i.e.} the momenta
that are connected by the AF ordering wavevector) and $Z_{\rm S}$,
the corresponding quasiparticle residue of the small Fermi
surface, are nonzero on the two sides of the QCP, but each one
approaches zero as the QCP is approached from the respective side.
Also shown are the illustrations of the large and small Fermi
surfaces, which refer to those with the \textit{f}-electrons
delocalized and localized, respectively. The actual Fermi surfaces
in both cases are multi-sheeted, and can be located in the first
Brillouin zone in a more complex way.}
\end{figure}

\newpage

\begin{figure}
\begin{center}
\includegraphics[width=\columnwidth]{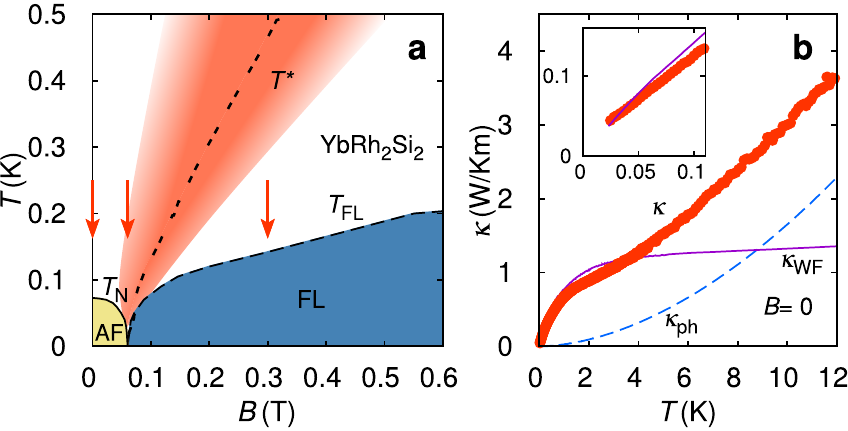}\\
\vspace{0.5em} {\Large Figure 1}
\end{center}
\end{figure}

\vfill
\begin{figure}
\begin{center}
\includegraphics[width=\columnwidth]{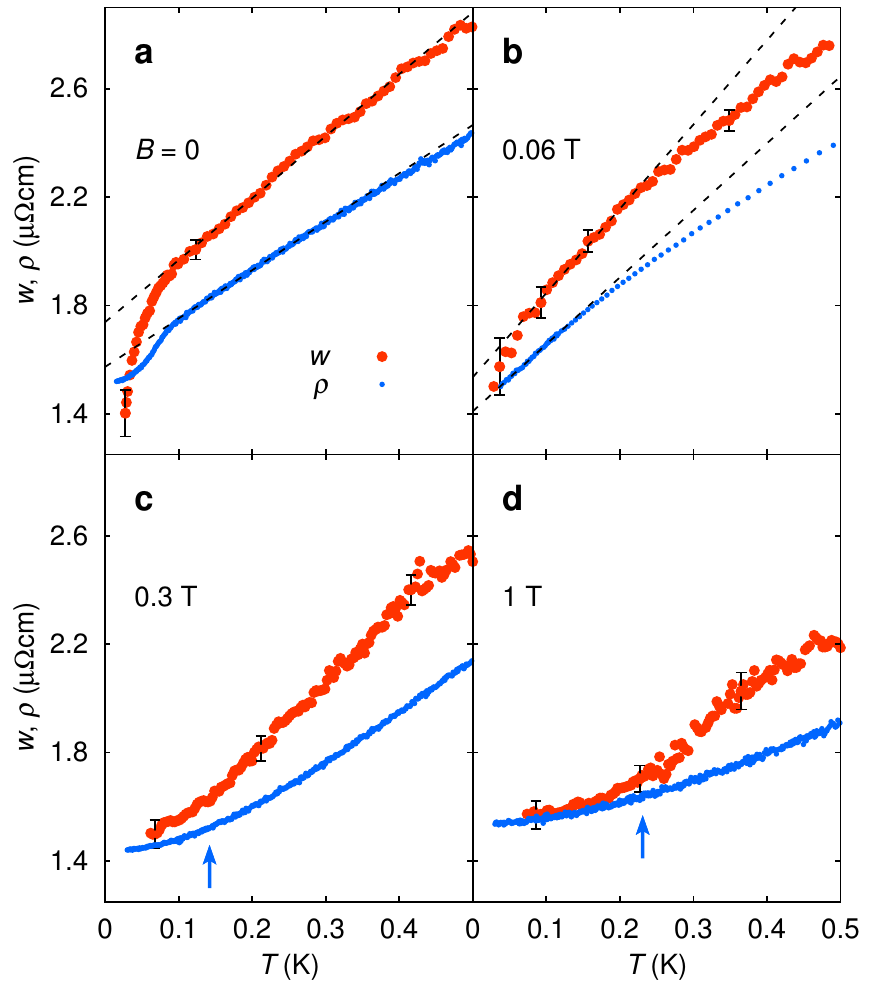}\\
\vspace{0.5em} {\Large Figure 2}
\end{center}
\end{figure}

\newpage
\begin{figure}
\begin{center}
\includegraphics[width=10.4cm]{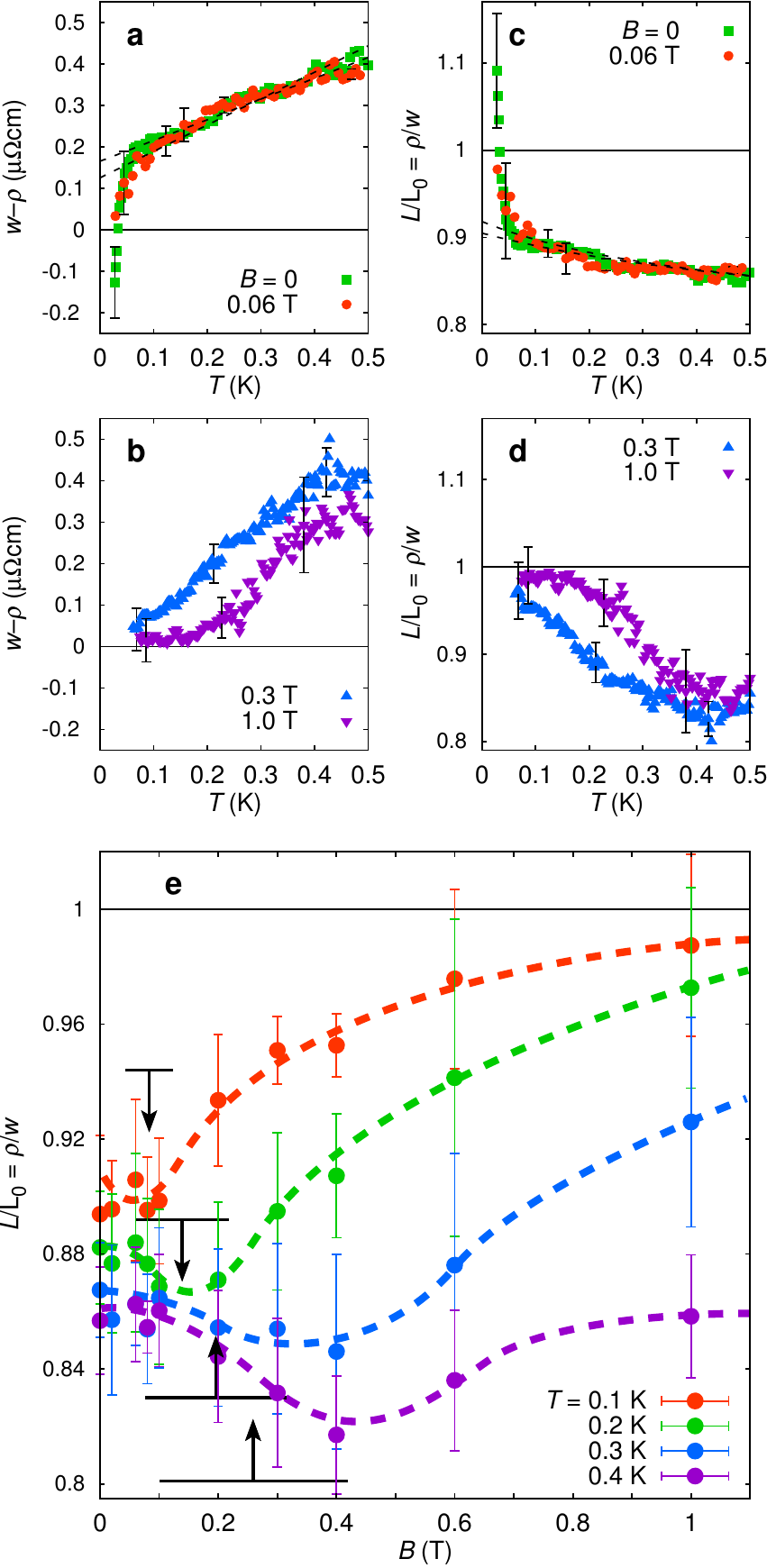}\\
\vspace{.5em}
{\Large Figure 3}
\end{center}
\end{figure}

\vfill
\newpage
\begin{figure}
\begin{center}
\includegraphics[width=10cm]{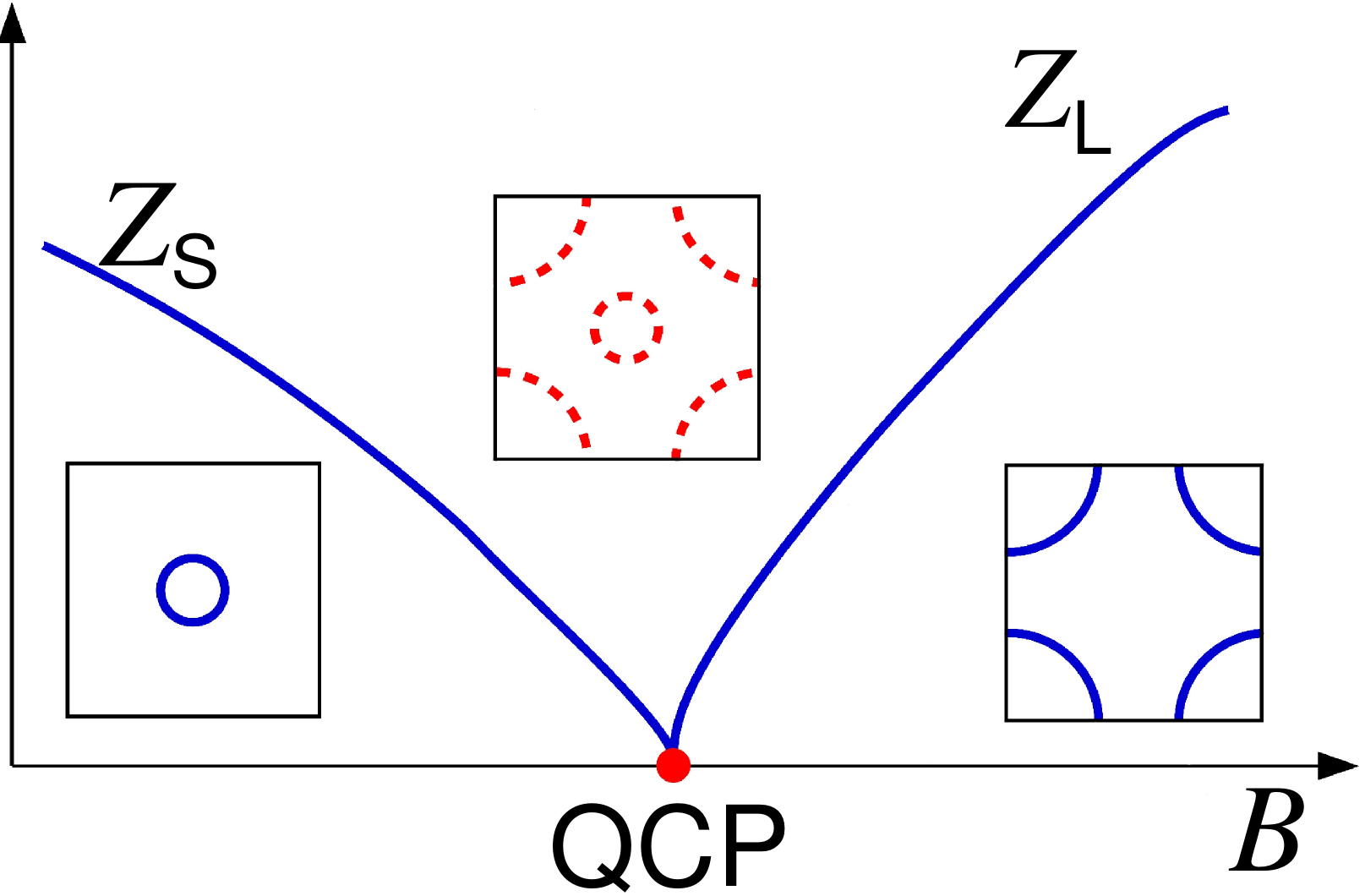}\\
\vspace{.5em}
{\Large Figure 4}
\end{center}
\end{figure}

\renewcommand\figurename{FIG. S$\!\!$}

\clearpage
\setcounter{figure}{0}
\title{Supplementary Information for\\
``Thermal and Electrical Transport across a Magnetic\\
Quantum Critical Point''}
%
%

\maketitle

\section{Background}
The low-temperature behaviour of the heavy-fermion metal \yrs was
studied by means of thermal and electrical transport across its
field-induced quantum critical point (QCP). The thermal
conductivity of metals is typically measured down to temperatures
in the Kelvin range.~\cite{2Smith08,2Wakeham11} Because of the low
energy scales, measurements in the heavy-fermion metals have been
extended down to temperatures as low as 0.06\,K
(Ref.~\onlinecite{2Tanatar07}) or even 0.04\,K
(Ref.~\onlinecite{2Tomokuni11}). In our case, special efforts have
been made to measure the thermal conductivity at $B = 0$ and
0.02\,T down to 0.025\,K due to the very low N\'eel temperature
$T_{\rm N}$ = 0.07\,K (at $B = 0$). Furthermore, we have performed
substantially more temperature scans for $B=0$ in order to reduce
the statistical error of the data. At $B = 0.06$\,T, which is
close to the critical field $B_{\rm c}$, we were able to measure
the thermal conductivity down to 0.04\,K. At higher fields, the
measurements of the thermal conductivity were performed down to
0.06\,K.
\section{Contacts}\label{sec:contact}
Previous thermal transport studies pointed out the strong
influence of thermal contact resistances and demonstrated the
necessity of a careful contact
preparation.\cite{2Tanatar07,2Seyfarth05} Subsequent to polishing
the \yrs sample (sample 1) for optimization of their geometry for
our measurements, the sample surface was cleaned in an ultrasonic
bath and rinsed in ethanol. The contact pads were prepared by
evaporating a gold film (thickness of $\approx 50$\,nm) and
applying a lift-off technique which allowed for optimum pad width
and separation. Gold evaporation was conducted by using electron
beam evaporators in ultra-high vacuum which turned out to provide
superior contacts compared to thermal evaporation. Gold wires
(50\,$\mu$m) were then attached to the contact pads by silver
paint to make use of the pad area. The silver paint covered an
area of width of about 130\,$\mu$m, which is very small compared
to the distance between the contacts (2.9\,mm). This contributes
to a systematic standard error of about 4.5\% in the estimation of
the geometry factor. It is the same for all experiments. We did
not observe any indication of diminished thermal conductivity due
to contact resistances down to 0.025\,K. The same contacts were
used for thermal and electrical conductivity measurements.
\section{Systematic errors}\label{sec:error}
Although most measurements were performed on a sample cut
specifically for thermal transport, sample 1 ($4.2 \times 0.5
\times 0.1$ mm$^{3}$), additional electrical resistivity
measurements were done on a second sample (sample 2) with average
size $1.7 \times 0.41 \times 0.06$ mm$^{3}$ from the same piece of
crystal. This was necessary to systematically study the dependency
of the low-$T$ electrical resistivity on the current, {\it cf.}
Section IV. Because of the inhomogeneous thickness of the samples,
the geometry factor could not be determined with very high
precision.
\begin{figure}[b]
\centering
\includegraphics[width=0.5\textwidth]{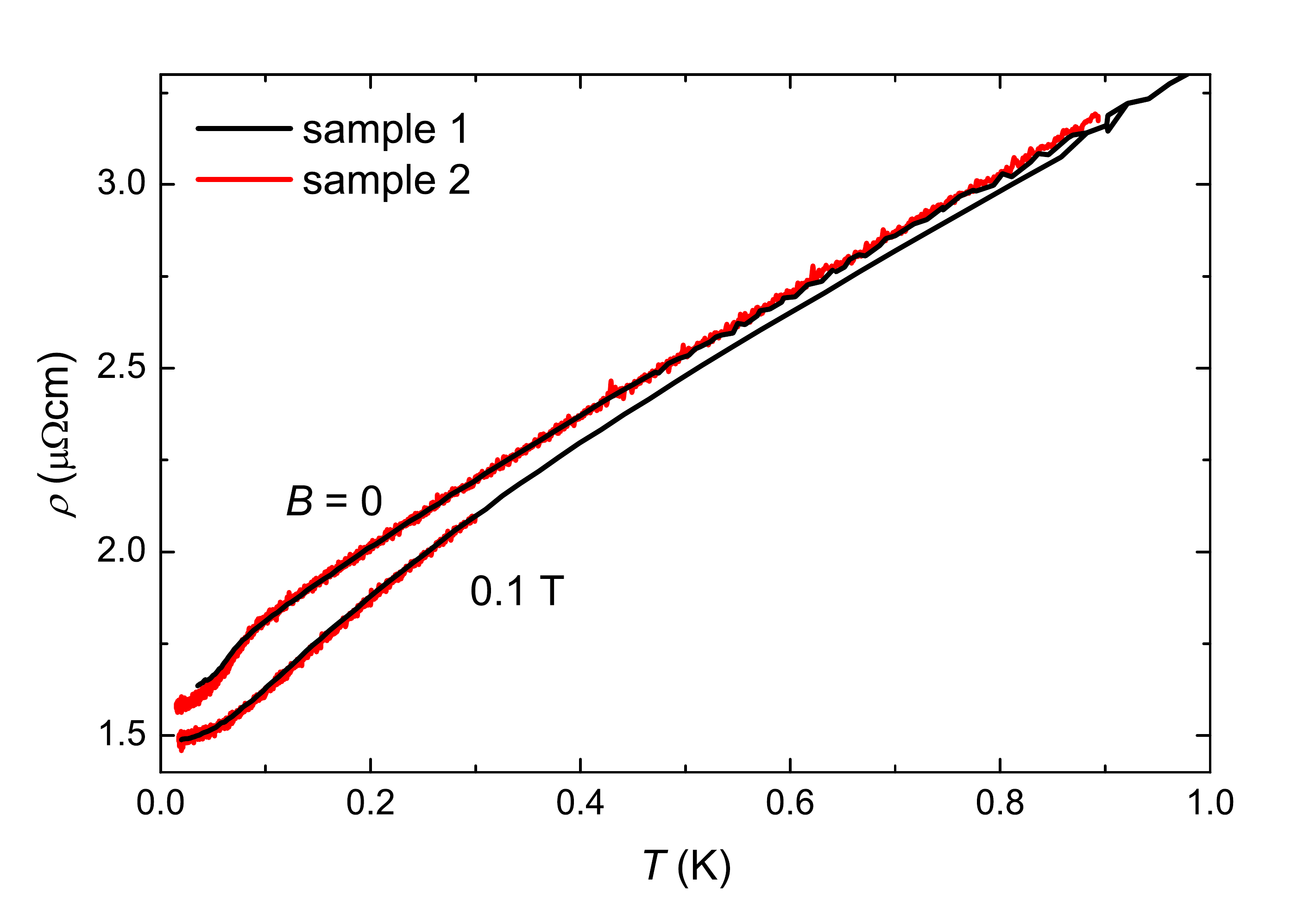}
\caption{Temperature dependency of the electrical resistivities of
two \yrs\,single crystals from the same batch.
The data of sample 2 have been multiplied by a
factor $1.25 \pm 0.03$ and corrected by a difference
in residual resistivity of 0.22\,$\mu\Omega$cm.}
\label{figS1}
\end{figure}
However, the measured resistivities could perfectly be rescaled by
a factor $1.25 \pm 0.03$ and corrected by a difference in residual
resistivity of 0.22\,$\mu\Omega$cm. The former number allows to
estimate the uncertainty in the determination of the sample
dimensions, {\it i.e.}, the systematic standard error for the
geometry factor. It is about 2.5\%, and it is the same for all
measurements. The thermal and electrical transport coefficients
were measured on the same sample (sample 1) with the same contact
configuration, and the Lorenz ratio $L(T)/L_0 =\rho(T)/w(T)$ is
affected by an additional systematic error of about 4.5\% due to
the finite width of the contacts, see Section~\ref{sec:contact}.
Thus, a total systematic error of 7\% has to be considered, {\it
i.e.}, it will shift systematically all curves of Fig.~3 of the
main text. This explains why in the region of the field-induced
Fermi-liquid in the phase diagram, {\it e.g.}, at $B \ge 0.6$\,T
and below $T$ = 0.15\,K, the difference $w(T) - \rho(T)$ is not
exactly zero, and the Lorenz ratio does not reach exactly one.
\section{Electrical resistivity}
The electrical resistivity $\rho(T)$ was determined by a
four-point ac-technique in a $^3$He-$^4$He dilution refrigerator
down to $T\approx$ 0.02\,K. Figure~S\ref{figS1} shows the
resistivity data of samples 1 and 2 at $B = 0$ and 0.1\,T,
respectively. The data for the second sample, which has a lower
residual resistivity, have been scaled as described above. The
data agree nicely, and we conclude that both samples show the same
overall behaviour. The slightly larger resistivity in the
\begin{figure}[t]
\centering
\includegraphics[width=0.5\textwidth]{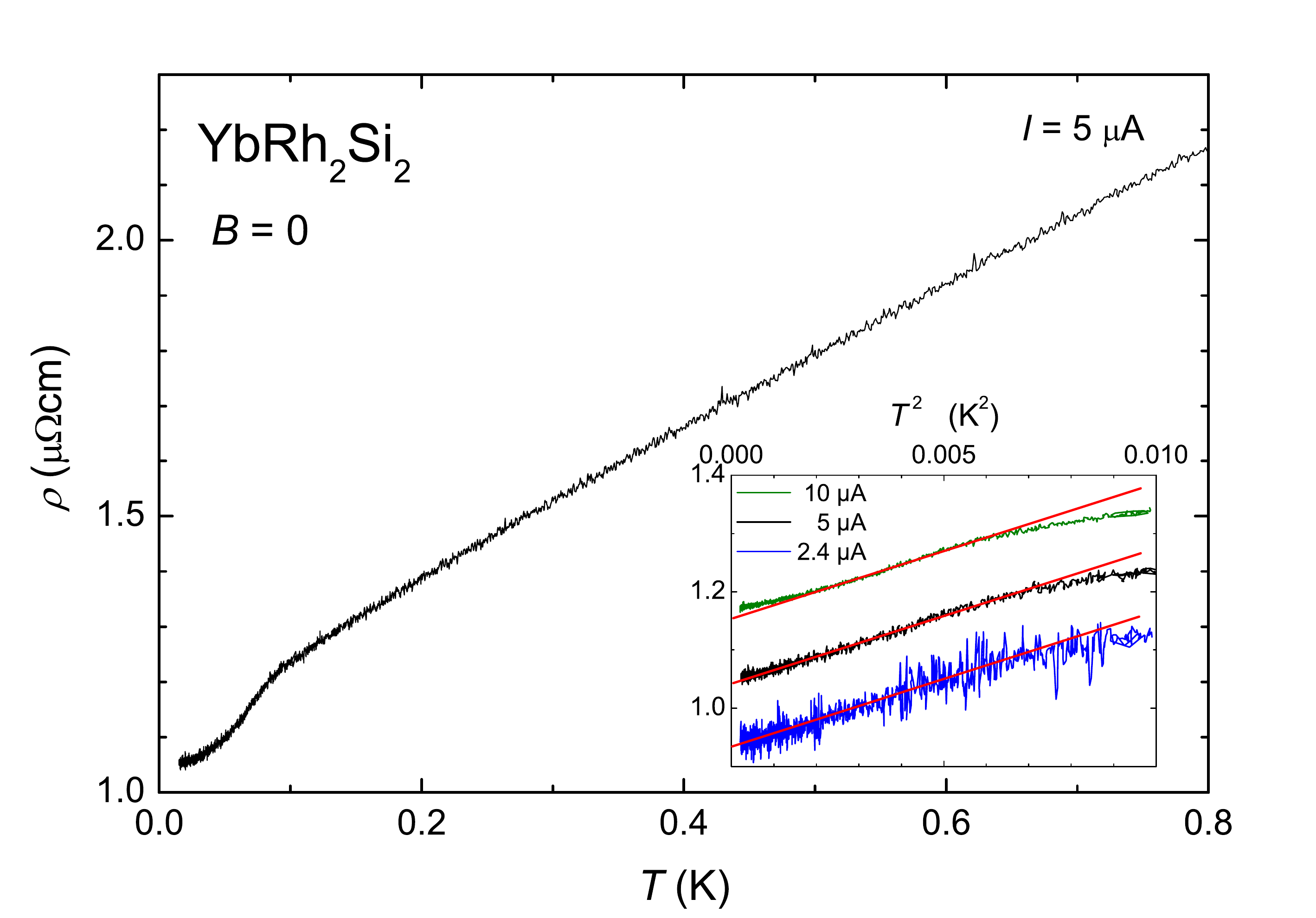}
\caption{Temperature dependency of the electrical resistivity of
sample 2. The kink at about 0.07\,K indicates the N\'eel
temperature \TN. Below \TN a $T^2$ dependency of the resistivity
is observed as expected~\cite{2Gegenwart02} ({\it cf.} inset where
the individual curves are shifted vertically).} 
\label{figS2}
\end{figure}
zero-field curve below 0.05\,K for sample 1 is due to heating
effects. Therefore, the second sample was measured at different
currents to investigate the influence of the current on the
curvature of $\rho(T)$. The results below 0.1\,K are displayed in
the inset of Fig.~S\ref{figS2} plotted as a function of $T^{2}$, a
temperature dependence expected below \TN = 0.07\,K for $B = 0$
(Ref.~\onlinecite{2Gegenwart02}). The heating effect can be
neglected at a current of $5\,\mu$A. We used the resistivity data
with this current to evaluate the Lorenz ratio. As will be shown
in a forthcoming paper,~\cite{2Lausberg} the heating effect is
particularly strong in the vicinity of $B_c$. In addition, it is
increasing with increasing residual resistivity $\rho_0$. For a
single crystal of very high perfection with $\rho_0 \approx 0.5
~\mu\Omega$\,cm, no heating effect could be observed down to $T
\approx 0.02$\,K using a current of 50\,$\mu$A
(Ref.~\onlinecite{2Westerkamp08}). In this case, $\Delta \rho \sim
T$ was observed below $T= 0.1$\,K down to the lowest accessible
temperature of 0.02\,K.

In all samples the magnetoresistivity was measured and compared to
that of the samples investigated in
Refs.~\onlinecite{2Friedemann10,2Friedemann11}, to assure consistency
of the properties observed in our samples with those of the best
single crystals grown so far. As an example, the
\begin{figure}[t]
\centering
\includegraphics[width=0.5\textwidth]{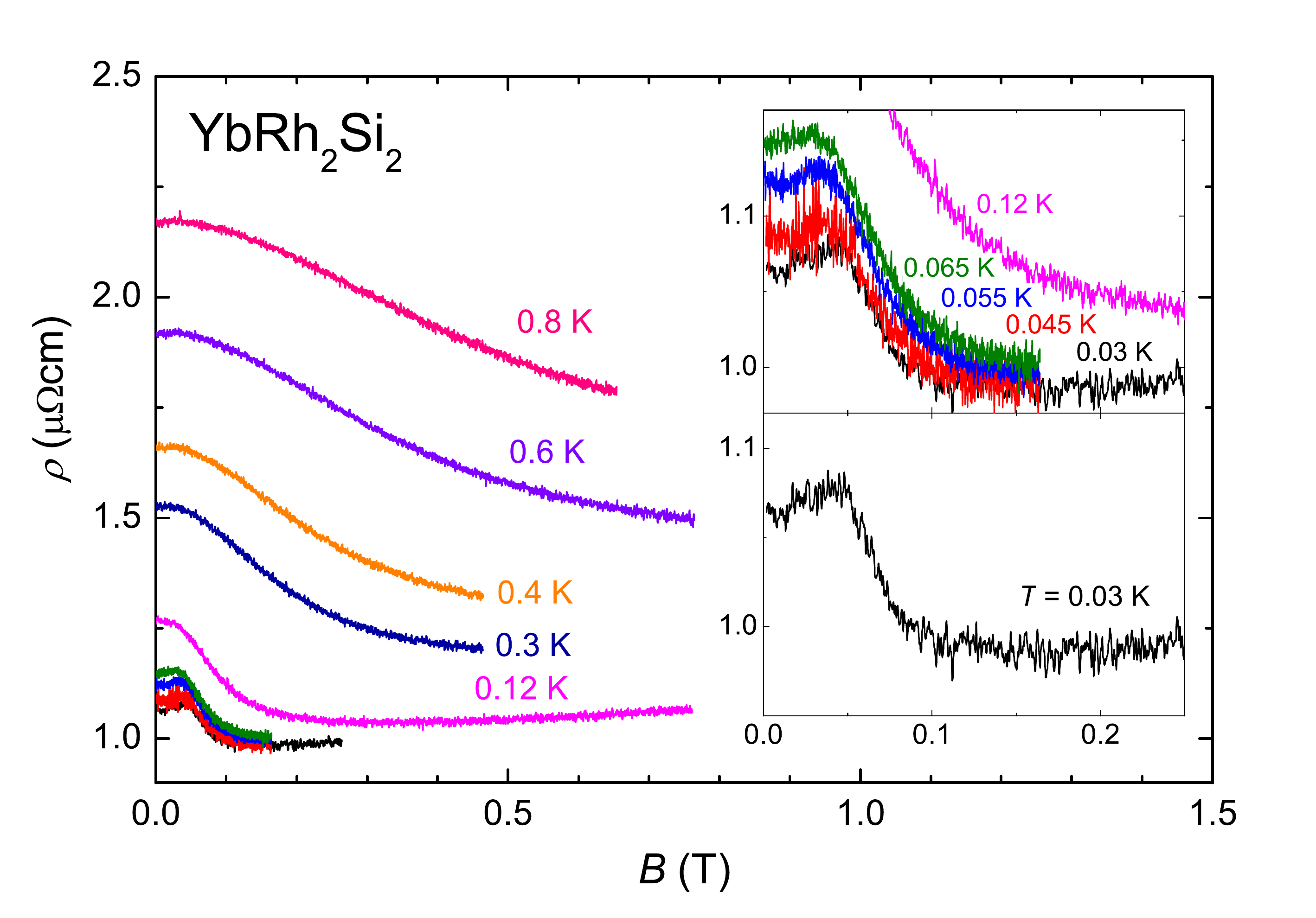}
\caption{Magnetoresistivity of sample 2 measured with a current of
$5\: \mu$A. The upper inset magnifies the low field, low
temperature region of the main panel to emphasize that the
magnetoresistivity is positive inside the AF phase, but exhibits a
rapid drop across a crossover field $B^{*}$. This is further
illustrated in the lower inset for our lowest temperature $T =$
0.03 K. The rapid crossover across $B^*$ was systematically
analysed \cite{2Friedemann10,2Friedemann11}; in the zero-temperature
limit, it corresponds to a sharp jump because the crossover width
extrapolates to zero. The low-temperature data are also consistent
with the existence of a peak in the residual resistivity as a
secondary feature superimposed on the sharp jump, although its
interpretation is not clearcut as it may also be associated with
the N\'eel transition. } 
\label{figS3}
\end{figure}
magnetoresistivity of sample 2 is shown in Fig.~S\ref{figS3}: At
$T = 0.03$\,K the change in resistivity
associated with the crossover is about 10\% (see insets of the
same figure), as also found in the samples studied in
Ref.~\onlinecite{2Friedemann10} ({\it cf.} Fig.~S5 of the
Supporting Information of Ref.~\onlinecite{2Friedemann10}). The
dominant feature is a rapid crossover across $B^*$, corresponding
to a jump in the zero-temperature limit
\cite{2Friedemann10,2Friedemann11}. The isothermal resistivity vs.
the magnetic field at low temperatures are also consistent with
the existence of a peak near $B^*$ as a secondary feature
superimposed on the sharp jump, although it could also be
associated with the (classical) N\'eel transition.

\section{Thermal conductivity}
\subsection{Experimental details}
\begin{figure}[t]
\centering
\includegraphics[width=0.47\textwidth]{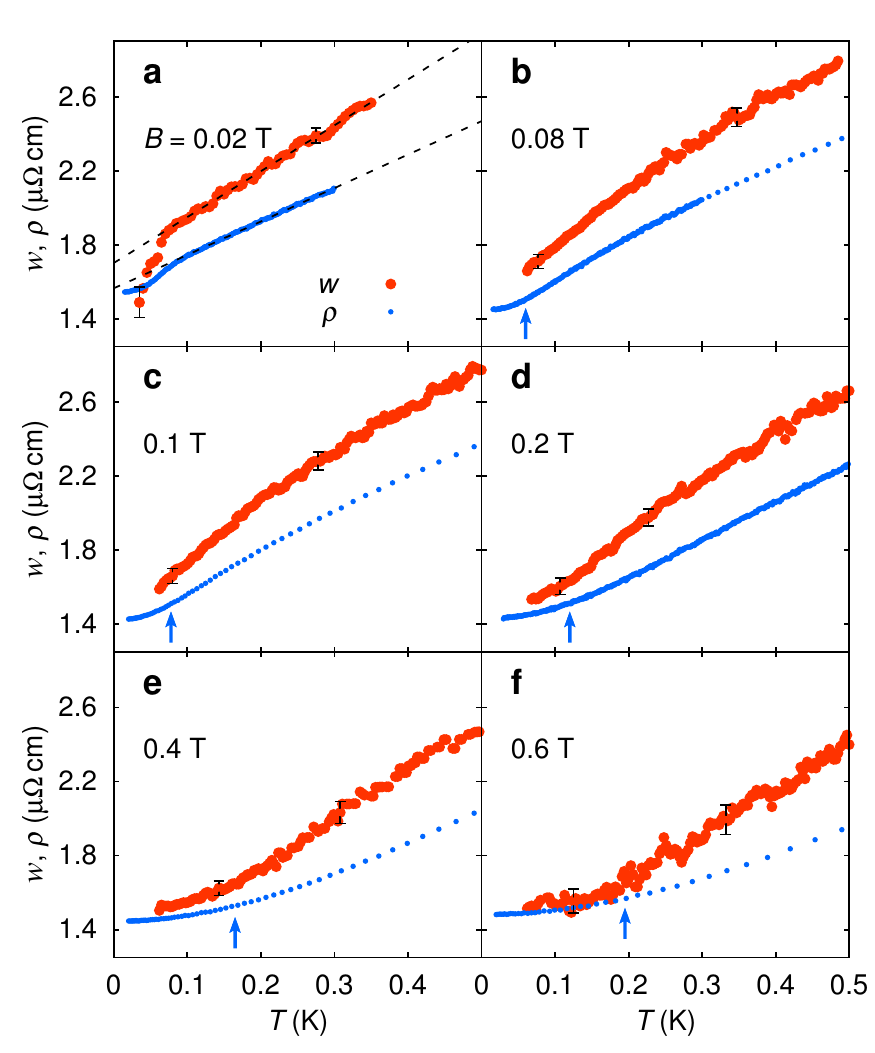}
\caption{Thermal and electrical resistivity $w = L_0 T/\kappa_{\rm
el}\approx L_0 T/\kappa$ and $\rho$ plotted as a function of
temperature, at $T \le 0.5$\,K for $B = 0.02$\,T
{\sffamily\bfseries a}, 0.08\,T {\sffamily\bfseries b}, 0.1\,T
{\sffamily\bfseries c}, 0.2\,T {\sffamily\bfseries d}, 0.4\,T
{\sffamily\bfseries e} and 0.6\,T {\sffamily\bfseries f} $(B \perp
c)$. The arrows in {\sffamily\bfseries b} through
{\sffamily\bfseries f} indicate the crossover to Fermi-liquid
behaviour ({\it cf.} Fig.~1{\sffamily\bfseries a} of the main
text). A significant drop is seen in the residual ($T\rightarrow
0$) thermal and electrical resistivities when going from $B =
0.02$\,T {\sffamily\bfseries a} to 0.08\,T {\sffamily\bfseries b}
and 0.1\,T {\sffamily\bfseries c}, which parallels the drop in the
isothermal magnetoresistivity across $B_{\rm c}$ ({\it cf.}
Fig.~S\ref{figS3}) as previously observed and attributed to an
abrupt increase in the charge carrier
concentration.~\cite{2Friedemann10,2Friedemann11} Also, the slight
rise of the residual thermal and electrical resistivities upon
increasing field ({\it cf.} {\sffamily\bfseries d} through
{\sffamily\bfseries f}) confirms the trend observed in
Ref.~\onlinecite{2Friedemann11} and ascribed there to the
magnetoresistivity in the paramagnetic phase of YbRh$_2$Si$_2$.}
\label{figS4}
\end{figure}

\begin{figure}[t]
\centering
\includegraphics[width=0.47\textwidth]{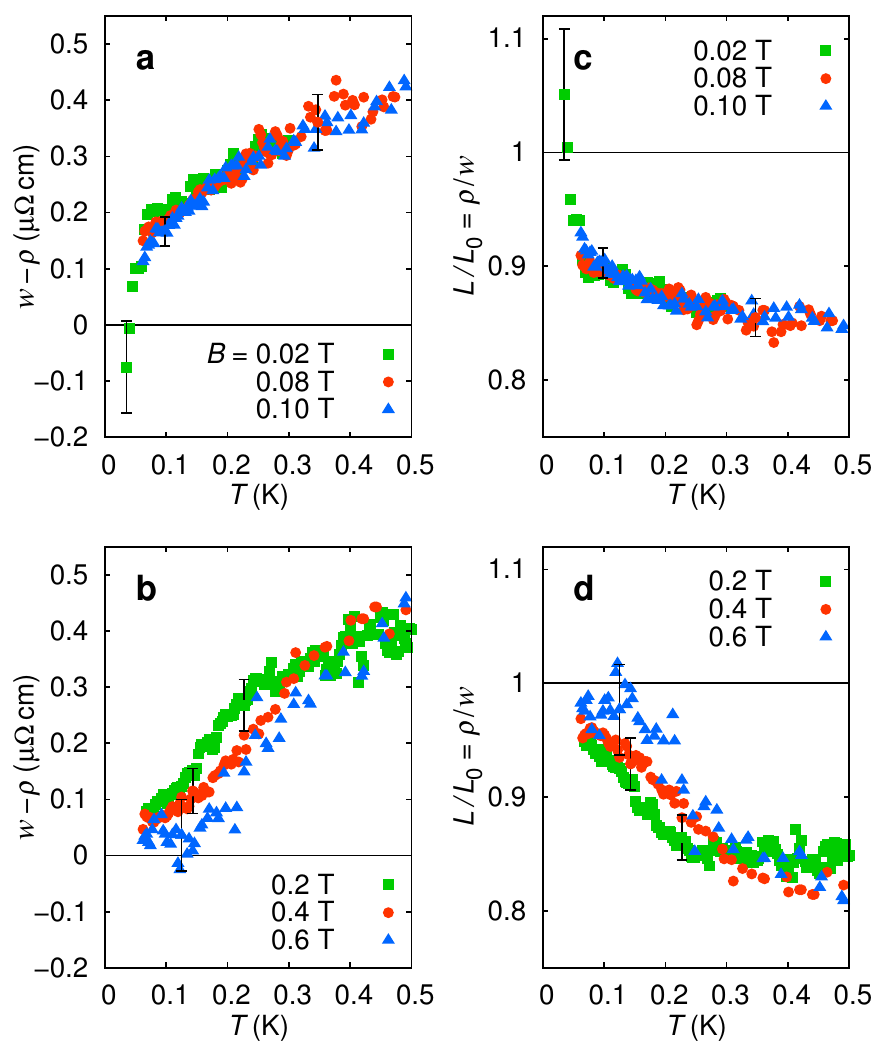}
\caption{Difference $(w - \rho)$ vs $T$ at $B = 0.02$\,T, 0.08\,T
and 0.1\,T {\sffamily\bfseries a} as well as 0.2\,T, 0.4\,T and
0.6\,T {\sffamily\bfseries b}. {\sffamily\bfseries c},
{\sffamily\bfseries d} Lorenz ratio $L/L_0 = \rho/w$ vs $T$ for
the same fields as in {\sffamily\bfseries a}, {\sffamily\bfseries
b}. The data for $B = 0.02$\,T are similar to those for $B = 0$
(\textit{cf.} Figs.~3{\sffamily\bfseries a} and
{\sffamily\bfseries c} of the main text). They provide convincing
evidence for an extra heat channel which adds to the one of the
electronic quasiparticles and ist most likely due to AF magnons
(see text). For $B = 0.6$\,T a Fermi-liquid phase forms below $T
\approx 0.15$\,K. This can also be anticipated for $B = 0.4$\,T
and 0.2\,T and even for $B = 0.1$\,T and 0.08\,T at
correspondingly lower crossover temperatures.} \label{figS5}
\end{figure}

The thermal conductivity $w(T)$ was measured using a steady-state
two-thermometer, one-heater technique in a $^3$He-$^4$He dilution
refrigerator down to $T\approx$ 0.025~K. The heat loss along the
wires and suspensions for thermometers and heater is estimated to
be a factor of 1000 smaller than the heat flow through the sample.
$\kappa$ was calculated from the electrical power $P$ released by
a resistive heater, the temperature difference $\Delta T$ between
two contacts on the sample and the geometry factor $A$ as $\kappa
= A \cdot P/ \Delta T$. $\Delta T$ was measured by two RuO$_{2}$
resistance thermometers, that were calibrated against the primary
(``cold finger``) sensor during each temperature run, using the
resistances measured at zero heating power. A stability of better
than 0.1\,\% of both sample temperature read-outs was achieved in
the entire measurement range. For each stabilized bath temperature
a set of four different heat currents was applied which results in
temperature gradients $\Delta T/T$ = 1\% - 7\% along the sample.
The proportionality between the applied heater power and the
achieved temperature gradients at a constant bath temperature
proves that the system is in the regime of linear response. The
resulting raw data of $\kappa$ were then averaged. Our measurement
procedure implies the following uncertainties: i) The uncertainty
in $A$ represents a systematic error, that shifts all $\kappa (T)$
curves by a constant factor, as described in
Section~\ref{sec:error}; ii) the uncertainty in $P$ is negligible
because current and voltage at the heater can be measured with
high accuracy; iii) the uncertainty in $\Delta T$ results from the
measurement of the thermometer resistances and from errors in the
calibration. Concerning the last point, two or more temperature
runs have been performed at many fields, each with its own
calibration. For a given field, the calculated $\kappa(T)$ values
fall on top of each other within the scattering of the data.
Therefore, the error in the calibration is negligible compared to
the data scattering. In fact, the largest uncertainty arises from
the measurement of the thermometer resistances due to the limited
excitation current, especially at low $T$. It leads to scattering
of the raw data and is significantly reduced in our final data by
averaging over several points. As error bars we took the standard
deviation of the raw data as displayed in Figs.~2 and 3 of the
main text as well as Figs.~S\ref{figS4} and S\ref{figS5}. The
systematic error due to the sample and contact geometry is not
included, because it results only in a shift of all curves as
explained in Section III. We attribute the observation of $w_0$
being about 1\% larger than $\rho_0$ in the Fermi liquid regime at
high fields (Figs.~2{\sffamily\bfseries d} and
S4{\sffamily\bfseries f}) to this systematic error.

Finally, the noise level in the data of the thermal resistivity
increases substantially with increasing magnetic field, as can be
directly seen in Fig.~S\ref{figS4}. It is presumably due to the
vibrations of the set-up wires in
magnetic fields. This is the main reason why we could not perform
reliable measurements below 0.06\,K at fields larger than 0.06\,T.
\subsection{Phonon contribution}
The phonon contribution, \kphT, to the measured thermal
conductivity $\kappa(T)$ can be separated from the electronic
part, \kelT, by avoiding the low-temperature range where a
significant inelastic scattering of the charge carriers has been
evidenced ({\it cf.} Fig.~1{\sffamily\bfseries b} of the main text
and Ref.~\onlinecite{2Tomokuni11}).

After having subtracted \kelT = $\kappa_{WF}(T) = L_0 T/\rho(T)$
from $\kappa(T)$ for 6\,K $< T < 12$\,K, {\it i.e.}, assuming the
validity of the Wiedemann-Franz (WF) law to hold in this
temperature range for the electronic heat transport, the phonon
contribution \kphT is found to follow a $T^{\epsilon}$ dependence
with $\epsilon = 2 \pm 0.2$. As shown in Fig.~1{\sffamily\bfseries
b}, main text, an extrapolation of this power law to lower
temperatures indicates a negligible \kphT below 1\,K, {\it i.e.},
within the temperature range of interest in the present work. The
uncertainty in the exponent has only little effect on the
estimated \kphT below 5\,K. The insignificance of \kphT below 1\,K
is further corroborated by the almost constant Lorenz ratio $L /
L_0 \approx 0.86$ within
the range $0.5$\,K $< T  < 1$\,K, presumably associated with a
measured thermal conductivity of purely electronic origin.
Assuming this Lorenz ratio to extend to above 2\,K, \kelT = $L\,
T/\rho(T)$ can easily be estimated. Subtracting this from the
measured $\kappa(T)$, the phonon part is obtained once more. Below
about 5\,K, \kphT obtained by the latter procedure is found to be
in very good agreement with that from the former one.

We, therefore, conclude that  below about 10\,K, \kphT of \yrs
follows a $T^{2}$ dependence, as commonly expected for a metal in
this temperature range due to dominating phonon scattering from
the conduction electrons.~\cite{2Ziman60} A nearly
$T^{2}$-dependence of \kphT has indeed been observed for several
heavy-fermion compounds, e.g., CeB$_{6}$,~\cite{2Peysson85}
CeNiSn~\cite{2Kitagawa02} and CeAuAl$_{3}$~\cite{2Aoki00} in a
similar temperature range. In these systems the dominant phonon
wavelength, $\lambda_{\rm ph}$, is assumed to be shorter than the
mean free path of charge carriers, $l_{\rm el}$,
Ref.~\onlinecite{2Zimmerman59}. This assumption, thus, appears to
hold also for the \yrs single crystal studied here. On the other
hand, for CeCu$_{2}$Si$_{2}$,~\cite{2Franz78}
CeAl$_{3}$~\cite{2Ott84} and CeCu$_{6}$~\cite{2Peysson86} \kphT
depends almost linearly on $T$, which hints at the opposite
relation between $\lambda_{\rm ph}$ and $l_{\rm el}$, {\it i.e.},
$\lambda_{\rm ph} > l_{\rm el}$, Ref.~\onlinecite{2Zimmerman59}.
\subsection{Electronic contribution}
\begin{figure}[t]
\centering
\includegraphics[width=8.4cm]{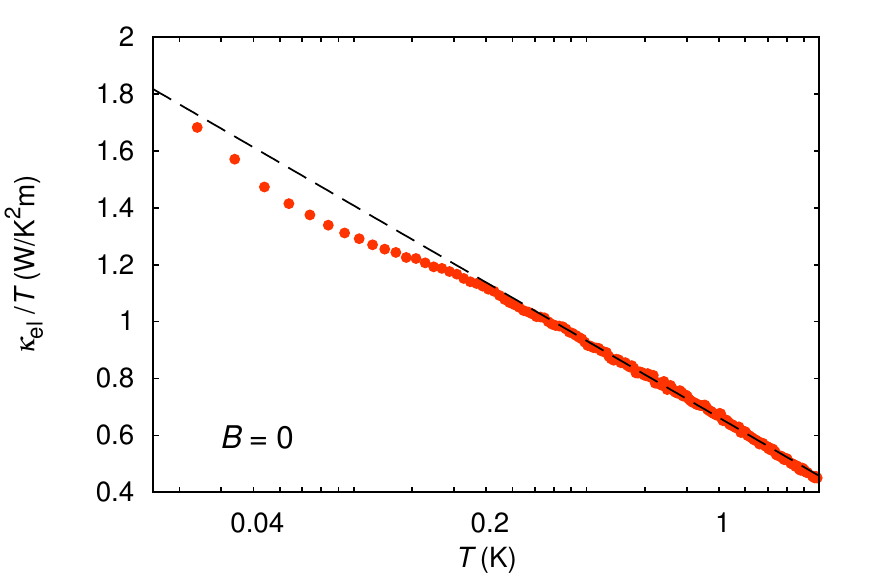}
\caption{Temperature coefficient of the $B = 0$ electronic thermal
conductivity, $\kappa_{\rm el}/T$, of \yrs plotted as a function
of $T$ on a logarithmic scale to emphasize that it follows a
logarithmic behaviour on cooling from 2\,K to about 0.3\,K.}
\label{figS6}
\end{figure}

The thermal conductivity obtained by subtracting \kphT from the
measured $\kappa(T)$ is $\kappa_{\rm el}(T)$. In Fig.~S\ref{figS6}
we show that the temperature coefficient, $\kappa_{\rm el}/T$,
increases logarithmically upon cooling from $T = 2$\,K to about
0.3\,K. A corresponding logarithmic divergence was observed also
for the Sommerfeld coefficient of the electronic specific heat,
$\gamma = C_{\rm el}/T$,~\cite{2Trovarelli00,2Custers03} and the
thermopower coefficient $S/T$.~\cite{2Hartmann10}

In the following, we are interested in the corresponding thermal
resistivity $w(T) = L_0 T/\kappa_{\rm el} (T)$, which is displayed
together with $\rho(T)$, in Fig.~2 of the main text for four
different fields. The results of $w(T)$ and $\rho(T)$ for other
magnetic fields are shown in Fig.~S\ref{figS4}. The data taken at
$B = 0.08$\,T through $0.6$\,T are complementary to those at $B =
0.3$\,T and 1 T presented in Figs.~2{\sffamily\bfseries c} and
2{\sffamily\bfseries d} of the main text. The data at $B =
0.02$\,T are likewise complementary to those at zero field
presented in Fig.~2{\sffamily\bfseries a} of the main text. Like
in the zero-field case, they show a low-temperature downturn in
the AF phase which contains the magnon contribution to the heat
conduction. Figure S\ref{figS5} displays the difference $w(T) -
\rho(T)$ {\sffamily\bfseries a},{\sffamily\bfseries b} and the
ratio $\rho(T)/w(T) = L/L_0$ {\sffamily\bfseries
c},{\sffamily\bfseries d} for the data shown in Fig.~S\ref{figS4}.
The existence of a magnon contribution to the thermal conductivity
at $B = 0.02$\,T is indicated by $(w - \rho)$ being negative and
$L/L_0$ being larger than one below $T \approx 0.03$\,K, very
similar to the results at $B = 0$ (\textit{cf.}
Figs.~3{\sffamily\bfseries a} and 3{\sffamily\bfseries c} of the
main text). Within the experimental uncertainties, the data at $B
= 0.6$\,T indicate Fermi-liquid behaviour and the validity of the
WF law below $T \approx 0.15$\,K. Fermi-liquid behaviour below a
crossover temperature which continuously decreases with decreasing
magnetic field is inferred for both $B = 0.4$\,T and 0.2\,T
{\sffamily\bfseries b},{\sffamily\bfseries d}. This implies
$L/$L$_{0} \rightarrow 1$ as $T \rightarrow 0$, with which also
the results obtained for $B = 0.1$\,T and even 0.08\,T are
compatible {\sffamily\bfseries a},{\sffamily\bfseries c}.

The data for $w$ and $\rho$ measured as a function of temperature
for fixed magnetic fields, shown in Fig.~2 of the main text and
Fig.~S\ref{figS4}, have been used to determine the isothermal
Lorenz ratio as a function of field displayed in
Fig.~3{\sffamily\bfseries e} of the main text. The isothermal
field dependence exhibits a minimum near $B^*$. This striking
behaviour is also seen in preliminary results obtained from direct
isothermal measurements of $w$ and $\rho$ as a function of field
in a sample with lower residual resistivity~\cite{2isothermal_new}.

\subsection{Magnon contribution}
For the measurements performed at $B = 0$
(Fig.~2{\sffamily\bfseries a}, main text) as well as 0.02\,T
(Fig.~S\ref{figS4}{\sffamily\bfseries a}) a downturn in the
thermal resistivity is found below $T \approx 0.1$\,K and $T
\approx 0.08$\,K, respectively, while at $B = 0.06$\,T a similar
but smaller feature occurs below $T \approx 0.07$\,K
(Fig.~2{\sffamily\bfseries b}, main text). One may ascribe this
drop in $w(T)$ to the freezing out of inelastic scatterings
provided by spin fluctuations in the electronic heat transport.
However, while ferromagnetic spin fluctuations remain unchanged,
antiferromagnetic ones grow at sufficiently low temperatures and
magnetic fields, as inferred from NMR Knight shift and
spin-lattice relaxation rate measurements, respectively.
\cite{2Ishida.02} Therefore, the only natural interpretation of
this drop involves heat carriers which add to the electronic ones.
The existence of a corresponding excess contribution to the
thermal conductivity, $\Delta \kappa (T)$, is proven by our
observation that for $B=0$ and $B=0.02$ T the thermal resistivity
becomes smaller than the electrical resistivity, {\it cf.}
Figs.~2{\sffamily\bfseries a} and S4{\sffamily\bfseries a}.
$\Delta \kappa (T)$ adds to $\kappa_{\rm el} (T)$ which we assume
to be given by $L_0 T/\rho$ at temperatures well below \TN, {\it
i.e.}, we assume the Wiedemann-Franz law to describe the
electronic transport in the Fermi-liquid phase well below $T_{\rm
N}$. The observed $\Delta\kappa(T)$ falls into the range $2 -
5\cdot10^{-3}$\,W/Km between 0.025 and 0.03\,K. Because this value
of $\Delta\kappa$ is close to the experimental uncertainty, its
temperature dependence is hard to be experimentally determined.
Nevertheless, its existence has been confirmed by repeated
measurements. Lattice vibrations are unapt to account for this
extra thermal conductivity, as the largest expected \kph, which is
limited by the sample dimension ($\approx 100\, \mu$m) and,
employing the lattice specific heat,~\cite{2Custers03} is estimated
to be less than $1\cdot 10^{-5}$\,W/Km in this range.

Another potential thermal heat channel is that of AF magnons. The
signature of these spin-wave excitations was clearly observed
\begin{figure}[t]
\centering
\includegraphics[width=0.52\textwidth]{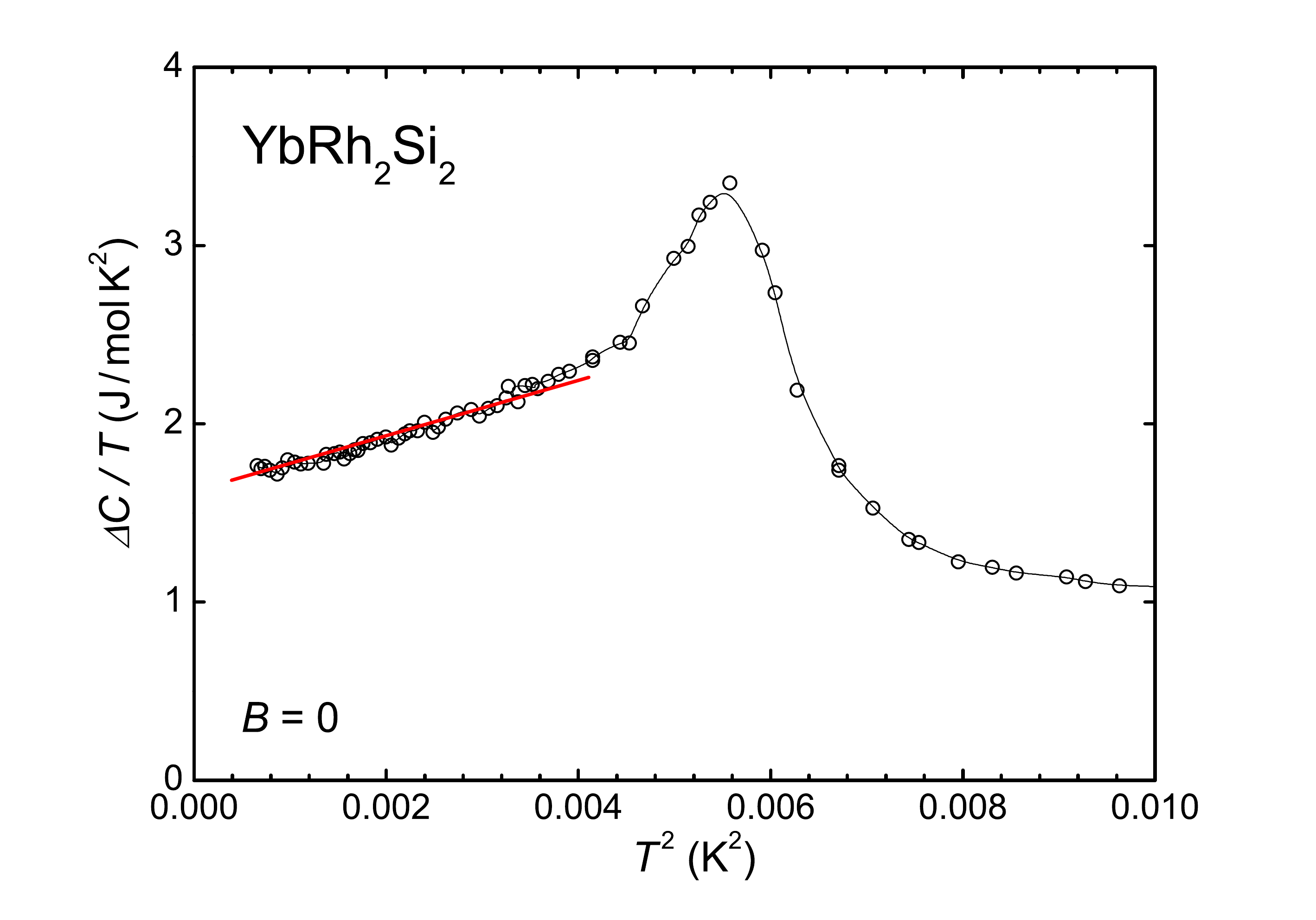}
\caption{Specific heat of \yrs, shown as $\Delta C /T$ versus
$T^{2}$. $\Delta C(T) = C(T) - C_{\rm ph}(T) - C_{\rm Q} (T)$,
where $C_{\rm ph}$/$C_{\rm Q}$ denotes the phonon/nuclear
quadrupole contribution. The red line indicates a $T^{3}$
contribution to $\Delta C$ below 0.05\,K.~\cite{2Custers03}}
\label{figS7}
\end{figure}
\cite{2Custers03} in the zero-field specific heat examined down to
$T = 0.018$\,K for an \yrs single crystal being of similar quality
as samples 1 and 2. As seen in Fig.~S\ref{figS7}, below $T =
0.05$\,K the specific heat can be described by $\Delta C = C -
C_{\rm ph} - C_{\rm Q} = \gamma T + \beta T^{3}$, with $C_{\rm
ph}$ and $C_{\rm Q}$ being, respectively, the phonon and the
nuclear quadrupole contributions, $\gamma = 1.64$\,J/(K$^{2}$mol)
and $\beta = 132.2$\,J/(K$^{4}$mol). The huge electronic
contribution $C_{\rm el} = \gamma T$ denotes a heavy Landau FL
phase.~\cite{2Custers03} The Debye-like term $C_m \sim T^{3}$ is
characteristic of the contribution of long-wavelength AF acoustic
magnons. In the framework of the Debye theory and using the
measured $\beta$ value, we can estimate the ``magnetic Debye
temperature", $\Theta_m$, and the group velocity $v_{m}$ of the
magnons to be 4.2\,K and 36\,m/s, respectively. This Debye
temperature corresponds to the AF exchange interaction suitably
averaged over the three main spatial directions. The extracted
group velocity is substantially smaller than that for typical
Ce-based heavy-fermion AF metals, {\it e.g.}, about one-tenth of
the corresponding value for CeAl$_{2}$,~\cite{2Bredl78} reflecting
both the averaging over the spatial directions as well as the weak
N\'eel order of \yrs. The classical kinetic relation, $\kappa_{m}
= (1/3) C_{m} v_{m} l_{m}$ ($l_{m}$: magnon mean free path),
allows to estimate the magnon contribution to the thermal
conductivity, $\kappa_{m}$. In order to obtain a $\kappa_{m}$ of
$2 - 5\cdot10^{-3}$\,W/Km in the range $0.025 - 0.03$\,K as
observed in Fig.~2{\sffamily\bfseries a} of the main text and
Fig.~S\ref{figS4}{\sffamily\bfseries a}, the corresponding magnon
mean free path $l_{m}$ has to be in the range of $2 - 13$\,$\mu$m.
Since in the low-temperature limit $l_{m}$ is expected to be equal
in size to that of the AF domains, the latter are estimated to be
of the order of a few $\mu$m below about 0.03\,K which, indeed, is
a reasonable order of magnitude.~\cite{2Note}

The less pronounced downturn in $w(T)$ observed below $T =
0.07$\,K at $B = 0.06$\,T (Fig.~2{\sffamily\bfseries b}, main
text) is ascribed to overdamped AF magnons, which were shown, via
inelastic neutron scattering
experiments,\cite{2Wiltshire83,2Wiltshire85} to exist for
antiferromagnetically ordered materials substantially above the
N\'eel temperature. Heat transport by short-lived magnon
excitations has been reported, e.g., for the parent compound of
the 214 high-$T_{c}$ cuprates, La$_{2}$CuO$_{4}$, an $S = 1/2$, 2D
antiferromagnet with a N\'eel temperature $T_{\rm N} \approx
310$\,K.\cite{2Hess03} In particular, $S = 1$ chain systems, like
Y$_{2}$BaNiO$_{5}$ (Ref.~\onlinecite{2Kordonis05}) and
Ni$($C$_{2}$H$_{8}$N$_{2})_{2}$NO$_{2}($ClO$_{4})$ (NENP)
(Ref.~\onlinecite{2Sologubenko08}) have served as model systems in
this context.

We would like to note that the magnon contribution cannot be
avoided by performing thermal Hall measurements. In contrast to
the case of phonons, \cite{2Wakeham11} magnons may generate a
transverse thermal gradient. \cite{2Onose10}

\section{Extrapolation of the Lorenz ratio to zero temperature in
quantum critical systems}

We now discuss the various isofield and isothermal scans that are
used for the extrapolation of the Lorenz ratio in the vicinity o a
QCP. We will discuss the special case that is pertinent to \YRS,
namely a field-induced QCP separating an AF ordered phase and a
paramagnetic Fermi-liquid phase.

\begin{figure}[t]
\centering
\includegraphics[width=0.45\textwidth]{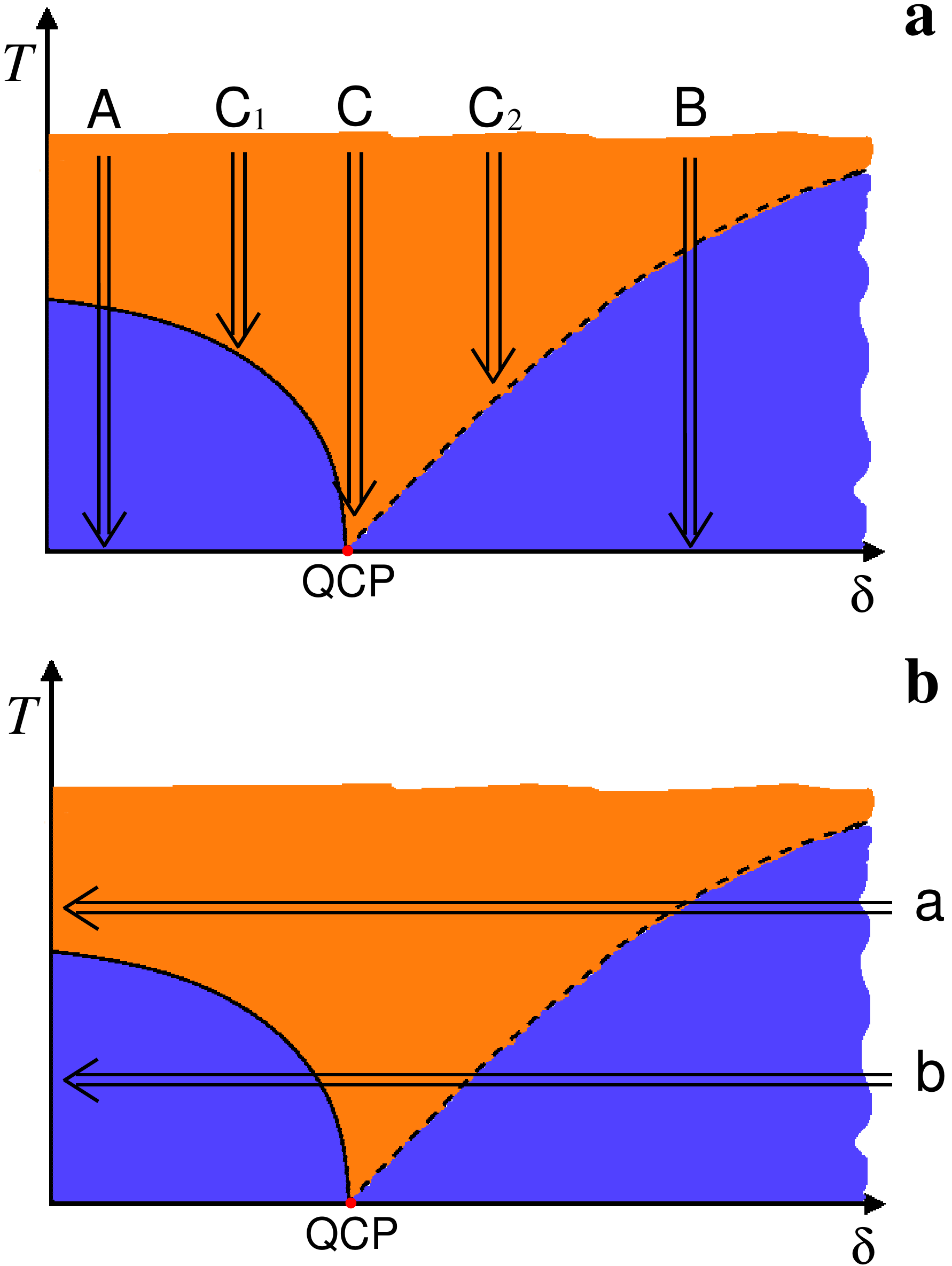}
\caption{ Isofield {\sffamily\bfseries a} and isothermal
{\sffamily\bfseries b} scans for the extrapolation of the Lorenz
ratio in the vicinity of a QCP, as described in detail in the
text. } \label{figS8}
\end{figure}

Fig.~S\ref{figS8}{\sffamily\bfseries a} illustrates different
temperature scans in different parts of the phase diagram. Scan A
starts from the quantum critical regime, but runs into the ordered
phase; by construction, it passes through the phase boundary. The
zero-temperature limit of $w(T)$, $\rho(T)$ and $L(T)/{L_0}$ so
extrapolated does not reflect the quantum critical behavior, but
instead only captures that of the ordered phase. Scan B also
starts from the quantum critical regime, but runs into the
low-temperature paramagnetic Fermi liquid phase; it passes through
a crossover temperature. Again, the zero-temperature limit so
extrapolated does not reflect the quantum critical behavior, but
instead only captures that of the paramagnetic Fermi liquid phase.
Scan C starts from the quantum critical regime, and goes all the
way to the QCP as the temperature is lowered. The zero-temperature
limit so extrapolated captures the properties of the QCP. In
practice, however, the same purpose can be achieved by carrying
out scans C$_1$ or C$_2$ (provided there is enough dynamical range
in temperature). These start in the quantum critical regime but
stop before running into the N{\' e}el temperature or the
crossover temperature to the paramagnetic Fermi liquid region. The
zero-temperature limit so extrapolated also manifests the behavior
of the QCP.

The different temperature scans will also be manifested in the
isothermal properties of $L/{L_0}$ as a function of the control
parameter, as illustrated in Fig.~S\ref{figS8}{\sffamily\bfseries
b}. Scan a starts from the paramagnetic Fermi-liquid regime, and
runs into the quantum critical regime. As it passes through the
crossover scale between the two regimes, we expect to see a
crossover in $L/{L_0}$ in a way that compliments what is seen in
the temperature scan B. Scan B starts from the paramagnetic
Fermi-liquid regime, passing through the quantum critical regime,
and runs into the ordered phase. Because it passes through both
the crossover temperature and the N\'eel temperature lines, this
isothermal scan will
reflect the features of both temperature scan B and temperature
scan A.

In this work, our primary focus is on temperature scans C$_1$
(Figs.\ 2{\sffamily\bfseries a}, 3{\sffamily\bfseries a},
3{\sffamily\bfseries c}), scans close to C (Fig.
2{\sffamily\bfseries b}), and scans C$_2$ ({\it e.g.} Figs.
S4{\sffamily\bfseries b} and {\sffamily\bfseries c}) through B
(Figs. 2{\sffamily\bfseries c}, 2{\sffamily\bfseries d},
3{\sffamily\bfseries b}, 3{\sffamily\bfseries d}). In addition,
isothermal scans a (Fig. 3{\sffamily\bfseries e}) are conducted.
Because of the masking effect of magnons on the electronic heat
transport, we do not attempt to analyze any scan b. Temperature
scans A cannot be performed in YbRh$_2$Si$_2$ because of the small
value of the critical field $B_c$.

\section{Theoretical aspects}
The validity of the WF law in the vicinity of a heavy-fermion
spin-density-wave QCP was briefly discussed in the main text.
Across such a QCP, the heavy quasiparticles remain intact in the
main part of the Fermi surface. In these ``cold'' regions, the
quasiparticles do not experience scattering by the AF spin
fluctuations, and their spectral weight---the quasiparticle
residue $Z$---stays finite as the control parameter moves through
the QCP (Fig.~S9). $L/{L_0}$ must be equal to one in this case.
Even the contributions from the ``hot'' regions, which experience
scatterings by the bosonic collective fluctuations, cannot violate
the WF law, as was illustrated by the case of ZrZn$_2$
(Ref.~\onlinecite{2Smith08}) where ferromagnetic fluctuations
influence the entire Fermi surface.

This is in contrast to the Kondo-destroying local quantum critical
description which is illustrated in Fig.~4 of the main text. In
the paramagnetic Fermi-liquid state, the conduction electron
\begin{figure}[t]
\centering
\includegraphics[width=0.45\textwidth]{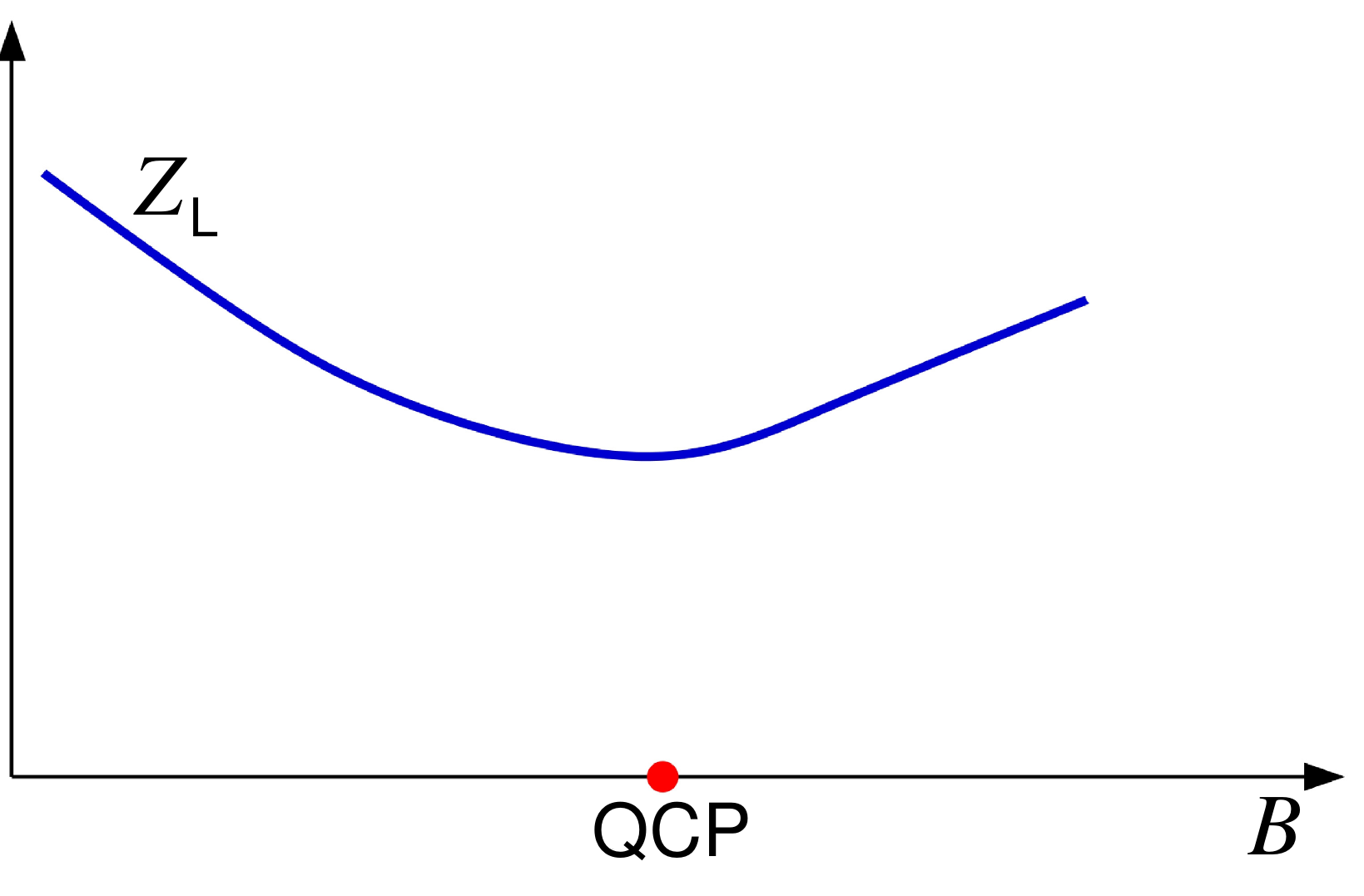}
\caption{The evolution of the quasiparticle weights across a
heavy-fermion spin-density-wave quantum critical points. $Z_L$ of
the large Fermi surface stays nonzero as the system is tuned
through the QCP.} \label{figS9}
\end{figure}
self-energy, $\Sigma ({\boldsymbol k},{\omega},T)$ contains a pole
at the Kondo-resonance energy. It converts the $f$-moments into a
part of the Landau quasiparticles, thereby creating a large Fermi
surface. The quasiparticle residue at the large Fermi momenta,
$\pmb{k}^{\rm L}_{\rm F}$, is nonzero, $Z_L(\pmb{k}^{\rm L}_{\rm
F},\omega=0,T=0) \ne 0$. In the AF Fermi-liquid state, the Kondo
resonance is destroyed. The Fermi surface is given by that of the
conduction electrons only in the presence of a staggered magnetic
field. Away from the ``hot spots", there are Landau quasiparticles
associated with a small Fermi surface. The quasiparticle residue
at such generic small Fermi momenta, $\pmb{k}^{\rm S}_{\rm F}$, is
nonzero, $Z_S(\pmb{k}^{\rm S}_{\rm F},\omega=0,T=0) \ne 0$.

\begin{figure}[b]
\centering
\includegraphics[width=0.45\textwidth]{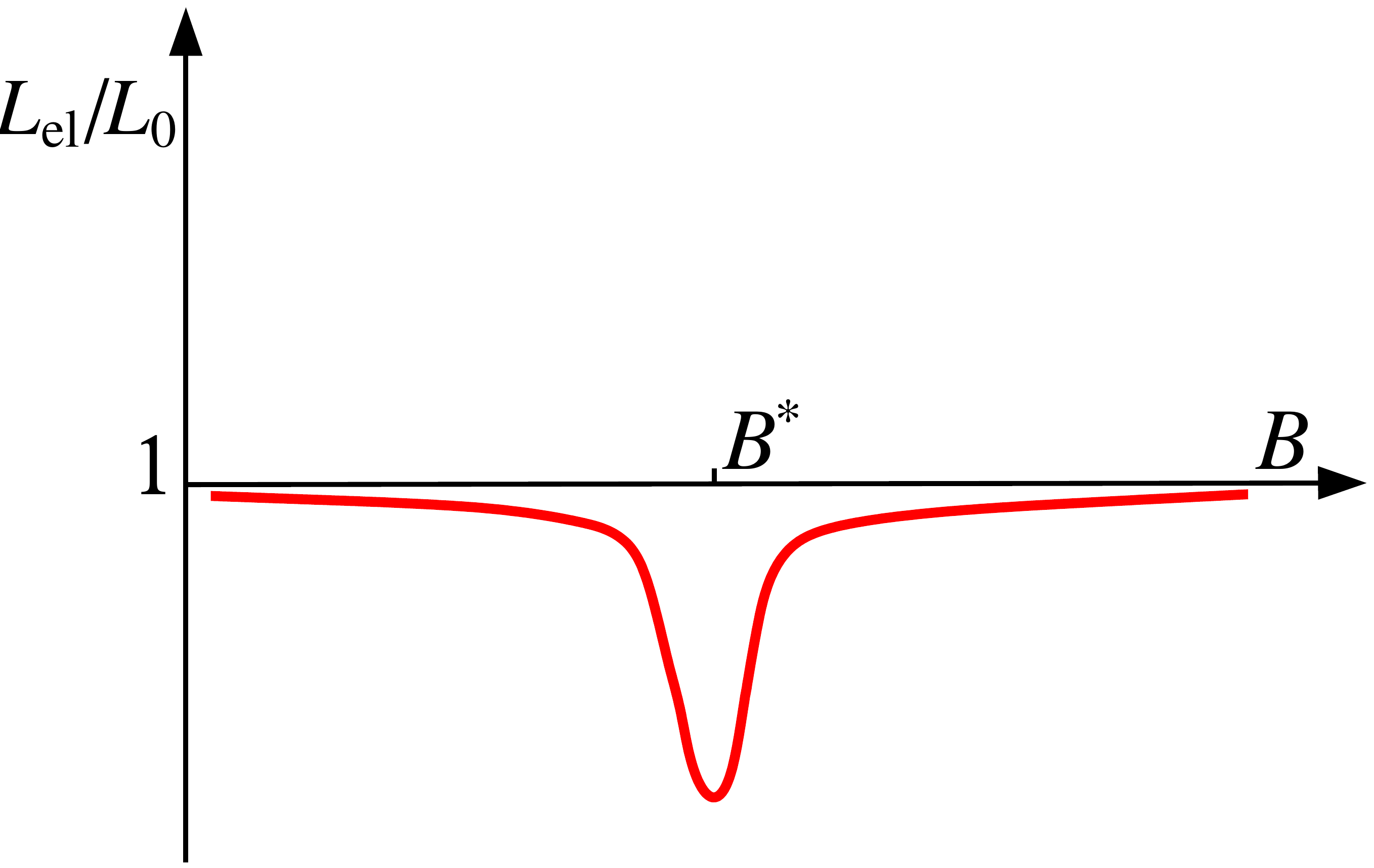}
\caption{Expected isothermal behaviour of the Lorenz ratio across
a Kondo-destroying QCP. $B^*(T) = B(T^*)$ is the Kondo breakdown
scale at a nonzero but small $T$. At zero temperature an abrupt
dip occurs at $B^* = B_c$} \label{figS10}
\end{figure}

In these Fermi-liquid regimes, both the electrical and electronic
heat currents are predominantly carried by Landau quasiparticles.
At nonzero temperatures, the electronic heat carriers experience
inelastic scatterings. Because the non-Umklapp processes are
considerably more efficient in relaxing electronic heat current
than electrical current, even in anisotropic systems at low
temperatures, the electron-electron scatterings lead to an
electronic Lorenz ratio $L_{\rm el}/{L_0} < 1$ at nonzero
temperatures.~\cite{2Herring67} The effect is similar to the usual
case of electron-phonon scattering.~\cite{2Ziman60} The Umklapp
scatterings are expected to contribute equally efficiently to the
thermal and electrical resistivities. In the zero-temperature
limit, well-defined quasiparticles remain, but the inelastic
scatterers are frozen out; only the elastic scattering processes
remain, and $L_{\rm el}/{L_0}$ is equal to $1$. This is
illustrated in Fig.~S\ref{figS10}, where $L_{\rm el}/{L_0}$
reaches $1$ as the system moves away from the QCP (at $B_c$) into
the Fermi-liquid regimes on both sides.

In the quantum critical regime, low-energy electronic excitations
occur at both the small and large Fermi surfaces (Fig.~4, main
text). The single-electron self-energy at both $\pmb{k}^{\rm
S}_{\rm F}$ and $\pmb{k}^{\rm L}_{\rm F}$ vanishes at the
$\omega=0$ and $T=0$ limit, but has the scaling form as a function
of $\omega$ and $T$ as given in Eq.~1 (main text). This reflects
the fluctuations of the Fermi surfaces, which
characterize the quantum fluctuations in the {\it entire} quantum
critical regime. We will consider the electrical and heat currents
carried by the electronic excitations at both $\pmb{k}^{\rm
S}_{\rm F}$ and $\pmb{k}^{\rm L}_{\rm F}$, all of which are
non-Fermi liquid in nature as specified by Eq.~1. These electronic
current carriers are subject to inelastic scatterings that are
associated with the quantum criticality. While such scattering
processes are many-body in nature, they can still be divided into
Umklapp and non-Umklapp processes. The non-Umklapp processes will
contribute considerably more to the thermal resistivity than to
the electrical resistivity. As a result, $L_{\rm el}/{L_0}$ will
be less than~1.

The end result is a dip of $L_{\rm el}/{L_0}$ near $B_c$.
Fig.~S\ref{figS10} illustrates the corresponding isothermal
behaviour of $L_{\rm el}/{L_0}$ for a given nonzero but low $T$,
where the dip is expected to be centered at $B^*(T) = B(T^*)$,
{\it cf.} Fig~1{\sffamily\bfseries a} of the main text. As $T
\rightarrow 0$, the dip becomes abrupt at $B_c$.


\end{document}